# Free-energy profiles along reduction pathways of MoS$_2$ M-edge and S-edge by dihydrogen: a first-principles study

Pierre-Yves Prodhomme (a), Pascal Raybaud (b), and Hervé Toulhoat (c)

(a) Direction Chimie et Physico-Chimie Appliquées, IFP Energies nouvelles, 1 et 4 avenue de Bois-Préau, 92852 Rueil-Malmaison, France,

(b) Direction Catalyse et Séparations, IFP Energies nouvelles, Rond-point de l'échangeur de Solaize, BP 3, 69360 Solaize, France

(c) Direction Scientifique, IFP Energies nouvelles, 1 et 4 avenue de Bois-Préau, 92852 Rueil-Malmaison, France

**Corresponding author:** Hervé Toulhoat,

IFP Energies nouvelles, 1 et 4 avenue de Bois-Préau, 92852 Rueil-Malmaison, France

Tel: +33.(0)1.47.52.73.50

Fax: +33.(0)1.47.52.70.22

e-mail: herve.toulhoat@ifpenergiesnouvelles.fr

**Abstract:**

We present the results of DFT calculations of free energy profiles along the reaction pathways starting from 50% coverage of $MoS_2$ M-edge and 100% coverage of S-edge by sulfur, and leading to 37% coverage, i.e. creation of anionic vacancies, upon reduction by dihydrogen and production of $H_2S$. Significant entropic and enthalpic corrections to electronic energies are deduced from the sets of normal modes vibration frequencies computed for all stationary and transition states. On that basis, we revisit and discuss the surface phase diagrams for M- and S- edges as a function of temperature, $H_2$ partial pressure and $H_2S/H_2$ molar ratio, with respect to ranges of conditions relevant to industrial hydrotreating operations. We show that in such conditions, anionic vacancies on the M-edge, and surface SH groups on the M- and S-edges, may coexist at equilibrium. Moderate activation barriers connect stationary states along all paths explored.

**Keywords :** Density Functional Theory, HDS catalysts activation, Free energy surface phase diagrams, Dihydrogen activation, Brönsted-Evans-Polanyi relationships, sulfur vacancies, SH and MoH surface groups vibrational frequencies

# 1    Introduction

To face an ever growing global energy demand, and depleting conventional oil resources, heavier crudes are increasingly refined into fuels. Simultaneously, local and global environment preservation regulations drive increased hydrogen addition, together with sulfur and other hetero-atoms removal, to end up with cleaner fuels. As such critical upgrades are achieved by catalytic hydroprocessing under a $H_2+H_2S$ atmosphere at high pressure, improved catalysts involving transition metal sulfides (such as $MoS_2$ based active phases) are keys to meet the rising challenges. However, basic questions such as the nature of actives sites and the mechanism of hydrogen activation remain so far much debated or unanswered [1, 2, 3, 4]. Providing well quantified data on the thermodynamic stability and activation pathways of hydrogen on $MoS_2$ is of paramount importance for a better understanding of hydrodesulfurization mechanisms. Indeed, determing the hydrogenation state of the catalyst in HDS working conditions will help for a better control of the surface chemical species: coordinatively unsaturated metallic sites (CUS), hydride, sulfhydryl group, sulfur anions. Most of these species are suspected to be involved in the elementary steps of HDS reactions [45]: reactant adsorption, monohydrogenation, hydrogenolysis, and sulfur vacancy regeneration.

Since most industrial hydroprocessing catalysts are based on $MoS_2$ nanoparticles, and it is now well established that catalytic activity is localized at their "M-edge" and "S-edge" surfaces, [4, 5] we establish here through first-principles (DFT) calculations including for the first time thermal and entropic effects on the catalyst, the free energy profiles of dihydrogen and hydrogen sulfide exchanges with such surfaces, including elementary steps of dissociative adsorption, associative desorption, and surface diffusion of bonded hydrogen. From previous works neglecting entropic contributions [3, 6, 7, 8, 9], no clear consensus emerges, regarding in particular the formation of sulfur vacancies and the nature and stability of hydrogen species at edges. For instance, Paul et al. [7] found that sulfur vacancy creations on S-edge and M-edge by $H_2$ to release $H_2S$, are very endothermic (1.31 eV for M-edge and 1.89 eV for S-edge), and kinetically difficult to achieve processes (1.47 eV for the "M-edge" and 2.10 eV for the "S-edge"). However, S vacancies were shown experimentally to be related with catalytic activity [10, 11], presumably offering chemisorption sites to hetero-aromatics and other unsaturated hydrocarbons and hence lowering activation barriers for further hydrogenation and hydrogenolysis steps, as moreover described by numerous DFT studies [1, 12, 13, 14]. Therefore, the present work aims at re-investigating the $H_2S$ associative desorption processes to assess not only the stability of S vacancies on $MoS_2$ edges but also the free energy barriers to vacancies formation.

Also, according to some of these works, H coverage was assumed to be hardly thermodynamically stable for few configurations [14, 8, 7], but since hydrogen adatoms on edges are expected to be primarily involved in hydrogenation and hydrogenolysis of adsorbed sulfided hydrocarbons, the reaction pathways for $H_2$ adsorption and H diffusion are re-investigated on both M- and S-edges.

In the present study, details on our DFT simulations are given in section 2.1. In order to assess the stability of the catalyst edges a thermodynamic model has been set up and is presented in section 2.2. Then, in section 3.1, are first presented the results of the computations performed on the reaction pathways for $H_2$ dissociative adsorption and $H_2S$ associative desorption along M-edge 50%S, S-edge 100%S, S-edge 87%S and S-edge 50%S. The surface phase diagrams for the S-edge and the M-edge derived from these pathways according to our thermodynamic model are then shown in section 3.2. In section 3.3 we compare the computed frequencies of normal modes of vibration for surface species to available experimental data, taking care to relate the conditions of measurement and preparation of the solids for the latter, to the predicted stable states according to our analysis. In this way, we did our best to look for crucial experimental evidence capable to invalidate our theoretical prediction. In section 3.4, we underline Brönsted-Evans-Polanyi Relationships (BEPR) emerging for the dissociative adsorption of $H_2$ and associative desorption of $H_2S$ on or from the surfaces considered in section 3.1. The discussion undertaken in section 4 focus on the one hand on the main differences between our phase diagrams and the former ones proposed [13, 15, 16], and on the other hand on the significance of our results with respect to the problem of activation by $H_2$ of $MoS_2$ based Hydrotreating catalysts  Finally, our conclusions are presented  in section 5.

## 2   Methods

### 2.1   DFT calculations

The computations were carried out according to the Density Functional Theory [17] with the Vienna Ab initio Simulation Package (VASP) [18], within the generalized gradient approximation, using the PW91 functional [19, 20], and the Projector Augmented Waves method [21] to treat the ionic-electronic interaction. The electronic density was calculated with a 350 eV cutoff energy, 3 k-points in the irreducible Brillouin zone [22], and a 0.1 eV electronic energy smearing.  For ground states the residual forces were set lower than 0.02 eV/Å. The Nudged Elastic Band (NEB) method was used to find the reaction paths [23], followed by the climbing image method [24], and the dimer method [25] to find the transition states (TS). The TS consistencies were verified when the higher forces applied on atoms were lower than 0.02 eV/Å, and the normal mode of the dynamical matrix exhibited only one imaginary frequency. In order to save computation time for the normal modes computation, only hydrogen, sulfur, and molybdenum atoms positions above the grey plane in figure 1 were kept free (corresponding to 20 atoms in figure 1). All other atoms were frozen. Normal mode frequencies were not corrected for anharmonicity.

In order to estimate the charge carried by each atoms a Bader analysis [26] was carried out on the electronic density computed with VASP (self-consistent calculations on the valence electronic density including 6 valence electrons per atom) thanks to the program established by Henkelman *et al.* from the University of Texas at Austin [27, 28, 29]. In this analysis, the net charge of one atom is the integral of the excess valence electronic density over the volume delimited by the zero-flux surface wrapping this atom minus 6e. On the zero-flux surface, the valence electronic density is minimal along the normal direction .

The parameters of the supercell, exhibited in figure 1, were similar to those previously used in ref. 3 and 42 but a Mo row was added in the x direction. Therefore, the model catalyst was composed of 4 Mo rows in the x direction 4 Mo rows in the y direction.  A 12.8 Å vacuum was allowed between each layer so that the slab was isolated in the y direction. A single Mo row separated by 12.3 Å vacuum was considered in the z direction, so that this layer was also isolated in this direction.

To find ground states, the simulated annealing method was applied, starting from the sulfur atoms on the edges in their bulk site. This heuristic method allows avoiding the system to become stranded in a local energy minimum. This simulated annealing procedure involved an ab initio molecular dynamics simulation on the Born-Oppenheimer surface, where the velocities were scaled every 5 steps, in order to reach after a while the targeted temperature. Between the rescaling steps the simulation was performed in the micro-canonical ensemble (constant N,V,E).

## 2.2 Thermodynamics corrections

The grand potential was computed in order to determine the most stable configurations. As the computed edges of the catalyst have all the same size, the grand potential was considered instead of the surface free energy:

$$\Omega\left(\mu_{M_o}, \mu_S, \mu_H\right) = E_{MoS_xH_y}^{tot+vib} - TS_{MoS_xH_y}^{vib} - n_{Mo}\mu_{Mo} - n_S\mu_S - n_H\mu_H \qquad [1]$$

where $E_{MoS_xH_y}^{tot+vib}$ is the total DFT energy plus the vibrational enthalpy of the $MoS_xH_y$ catalyst sheet containing $n_S$ sulfur, $n_{Mo}$ molybdenum, and with $n_H$ hydrogens on the edge. $S_{MoS_xH_y}^{vib}$ is the vibrational entropy of $MoS_xH_y$. $\mu_{Mo}$, $\mu_S$, $\mu_H$ are the chemical potentials of molybdenum, sulfur and hydrogen respectively. This accurate approach including thermal and entropic effects on the catalytic phase is provided for the first time on these systems to our knowledge. The enthalpies and entropies of vibration were computed thanks to the usual relationships derived from statistical thermodynamics [30], using the normal modes of the atoms near to the edge, and assuming that the normal modes of the atoms below the first two layers at the edge do not change. It is remarkable that the vibrational part of the free energy is not a constant term for the catalyst along the reaction path (see section 3). Hence for this reason the phase diagrams (figures 12, 14 and 15) have been computed for a set of temperatures representative of relevant HDS operations (473, 575 and 623 and 675 K).

As all the computed model catalysts contain the same number of Molybdenum atom, a curtailed grand potential is defined:

$$\Omega_c\left(\mu_S, \mu_H\right) = G_{MoS_xH_y} - n_S\mu_S - n_H\mu_H \quad [2]$$

where:

$$G_{MoS_xH_y} = E_{MoS_xH_y}^{tot+vib} - TS_{MoS_xH_y}^{vib} \qquad [3]$$

At thermodynamic equilibrium, the chemical potential of an atom is the same in all phases. So the chemical potential of sulfur and the chemical potential of hydrogen are determined within the gas phase [31]:

$$\mu_S = \mu_{H_2S} - \mu_{H_2} \qquad [4]$$

$$\mu_H = \frac{1}{2}\mu_{H_2} \qquad [5]$$

Total energies plus vibrational ground state energies for $H_2$ and $H_2S$ were computed at the DFT level, and the thermodynamic functions were calculated for the vibrational (null for $H_2$ and almost null for $H_2S$) and rotational parts of the chemical potentials of $H_2S$ and $H_2$. The translational parts were computed using the thermodynamical function in terms of the gas relative pressure [30]:

$$\mu_i^{trans} = -k_B T \ln\left(\frac{k_B T}{P_i}\left(\frac{2\pi m_i k_B T}{h^2}\right)^{3/2}\right) \qquad [6]$$

$P_i$, $m_i$, being the pressure and the mass of the particle $i$ respectively

The ideal gas approximation is obviously inadequate for high partial pressures in gas phase. Flash calculations on the basis of the Soave Redlich Kwong equation of state (SKR EOS) with volume translation [44] , give good estimates of the fugacities for $H_2$+$H_2S$ mixtures as function of temperature and pressure. They can be readily extended to account for the presence of liquid and gaseous hydrocarbons in the reactor. For instance at 200 bar total pressure, a mixture of 0.9 mole fraction $H_2$ and 0.1 mole fraction $H_2S$ corresponds to fugacities of 196 and 19.2 bar, 193.4 and 20.3 bar, 191.6 and 20.8 bar, for $H_2$ and $H_2S$ at 473K, 575K and 675K respectively. Equation 6 as well as phase diagrams of figures 10 and 11 remain valid in terms of fugacities. We recommend using the SKR EOS for the evaluation of the corresponding real partial pressures.

**3    Results**

In this section, we start by presenting in 3.1 total energy and free energy profiles along the reduction pathways of $MoS_2$ edges by dihydrogen, since the exploration of those pathways has been our strategy aiming at the determination of stable states according to catalytic operating conditions, in a similar way as our previous works ([3], [31], [32], [33]) and that of other authors ([2], [6], [7]). Having thus identified the local minima in free energy profiles, we construct in 3.2 surface phase diagrams allowing for the first time to identify the domains of prevalence of particular configurations at $MoS_2$ edges simultaneously as function of fugacities of $H_2$, $H_2S$, and of temperature, and therefore in reference to relevant operating conditions for catalytic hydrotreating. Further, in sub-section 3.3, we check our theoretical predictions against the results of spectroscopic experiments available from literature. This is done by comparing observed vibrational frequencies for MoS$_2$ based catalysts, prepared and analyzed in well defined conditions of temperature and fugacities of H$_2$ and H$_2$S, to our predicted frequencies assigned to surface groups normal modes, as expected for those conditions according to our surface phase diagrams. Finally, in section 3.4, we exhibit Brönsted-Evans-Polanyi Relationships for the H$_2$ dissociative adsorption and H$_2$S associative desorption processes on MoS$_2$ edge surface, as outcomes of the free energy profiles detailed in 3.1. From the latter, we therefore logically infer both the thermodynamic and kinetic consequences.

**3.1    Free energy profiles along reduction pathways of MoS$_2$ edges by dihydrogen**

This sub-section is to some extent the continuation of our previous work [3], in which we have presented the free energy profile along the partial reduction pathway starting from the Mo edge fully covered by sulfur dimers (M-edge 100%S) and leading to half-coverage by bridging S adatoms (M-edge 50%S). We therefore start with the reduction from M-edge 50%S to M-edge 37%S, which creates the first anionic vacancy in place of a bridging site. We verify then that the creation of the second anionic vacancy to give M-edge 25%S has a high free energy penalty so that this state will not appear. We proceed similarly for the S-edge, starting again from the full coverage in sulfur (S-edge 100%S), and ending with the thermodynamically lesser favoured S-edge 37%S after creation of 5 vacancies.

*3.1.1    M-edge (initial state 50%S)*

*3.1.1.1    Total energy profiles*

The heterolytic dissociative adsorption of a H$_2$ molecule is found to be an exothermic process (-0.12 eV), as shown in figure 2 (S1→S2). Moreover the energy barrier is found to be close to 0.48 eV. A careful analysis of the critical parameters of the calculation leading to these values in comparison with the open literature is reported in Supplementary Material for sake of clarity. In comparison the homolytic dissociative adsorption of $H_2$ on the M-edge with 100%S recently found by Dinter *et al.* [3] is less exothermic (- 0.08 eV) but the activation energy higher (0.95 eV). In this latter case, the physisorbed precursor is slightly more bound (-0.20 eV).

Once the dissociative adsorption occurred on the edge, the diffusion of hydrogen adatoms is kinetically favored compared with $H_2$ and $H_2S$ asssociative desorption, the latter having to overcome the largest barrier. Thus going through S2 and S3 states, diffusion leads to the most stable configuration (state S4_bis) of the reaction path, as shown in figure 2. This S4_bis state (see figure 3) is -1.09 eV more stable than the S1 state. It is reached by diffusion of H adatoms from the "trans" state S4, which is already -0.56 eV more stable than the S1 state (Sun et al. reported a stabilization by -0.37 eV for a closely similar configuration [8]). Notice that we find the "cis" S4 state destabilized by 0.19 eV with respect to "trans" S4, while the "cis" S4_bis state is -0.06 eV more stable than the "trans". The pathways from S4 to S4_bis pass through TS3_bis/S3_bis/TS3_ter. The barrier to TS3_bis is 0.68 eV, a little lower than the barrier from S4 to S3 via TS3 (0.70 eV). The intermediate S3_bis between S4 and S4_bis, is at -1.02 eV, almost as stable as S4_bis since the bridging S-H and the Mo-H are already separated by one bridging S. A further stabilization by -0.05 eV occurs with the diffusion of H along the next M-S bond, through transition state TS3_ter across a barrier of 0.31 eV. Here we should emphasize that an even longer distance between hydrogens of sulfhydryl groups might furthermore stabilize the hydrogenated edge. Because of the repulsive interaction between sulfhydryl groups, diffusion is essential to reach the most stable state on the edge. In addition, increasing hydrogen coverage beyond 2 H per 4 S atoms is not energetically favourable as also found in [8] (see also paragraph 3.2).

It is clear from figure 2 that hydrogens are more stable when bonded to sulfur than when bonded to molybdenum. This is not surprising as molybdenum atoms are already six-fold coordinated on the M-edge 50%S.

Starting from the S3 state the desorption barrier is 0.89 eV (close to the value reported by Paul et al. for a similar process [7], 1.0 eV), but the overall barrier is much larger if one considers desorption from the S4 state, 1.36 eV (0.83-(-0.53)), while the creation of a S vacancy is endothermic (+1.32 eV).

However, the overall reduction process starting with $H_2$ adsorption and ending up with $H_2S$ desorption is much more favoured thermodynamically ($\Delta E = 0.79$ eV ) and kinetically (barrier 0.83 eV) in comparison with Paul *et al.* results (1.31 eV and 1.47 eV respectively).

Here Fig. 2

Here Fig. 3

*3.1.1.2  Thermodynamic corrections*

In order to determine the stable configurations from a thermodynamic point of view, the thermodynamic contributions were included in figure 2 together with the total ground state DFT energy profile, giving estimates of the free energy profiles at 575K (blue line) and 675K (red line), under 10 bar of $H_2$ and 0.1 bar of $H_2S$ representative of a range of relevant HDS operating conditions.

The thermodynamic corrections slightly change the reaction path although the tendency remains similar, i.e. the S4_bis state is the most stable along the reaction path involving H diffusion (S2 to S5). Taking into account vibrational corrections decreases the activation energy for $H_2S$ desorption to 1.19 eV and 1.20 eV at 575 and 675 K respectively compared to 1.36 eV for DFT energy alone. Moreover, the extra difference brought by vibrational energy and entropy could be as much as 0.19 eV (between S3 and TS4) along the H diffusion path. The most drastic difference (0.37 eV) is found for the adsorption-desorption process (between S1 and S3). Such corrections lie well beyond the accepted accuracy of DFT.

The S6 state is, at equilibrium, almost as stable as the S1 state and its stability increases with temperature. This is interesting since this state involves an anion vacancy, which is an adsorption site for sulphided molecules like thiophenic homologues [32, 33]. This point will be discussed further in paragraph 3.2.

### 3.1.2 S-edge

#### 3.1.2.1 Total energy profiles

##### a) initial state S-edge 100% S

A quick simulated annealing from 900K down to 100K afforded us to find a stable configuration for the S-edge, where a "dimerized" $S_2$ bridge appears, see table 1 second row. This "dimerized" $S_2$ bridge appears to be more stable than the "separated" $S_2$ bridge. Several configurations have been computed with several "dimerized" $S_2$ bridges, all reported in table 1: four "dimerized" $S_2$ bridges (in this configuration the dimers bend on one side or the other alternately as mentioned by Bollinger et al. [15]),
two $S_2$ dimers (with two possible configurations), and one $S_2$ dimer on the edge. The
alternation of $S_2$ "dimerized" and "separated" $S_2$ bridges, further denoted as S1, is the
most stable configuration (by -0.99 eV) as proposed by Hinnemann *et al.* [34], even
more stable than four $S_2$ separated bridges. The change in vibrational normal modes
between no "dimerized" bridge and one "dimerized" bridge on the edge reduces this
energy difference by only at most 0.17 eV in free energy (at 675 K), whereas this
reduction is almost null between one "dimerized" bridge and two "dimerized" bridges
on the edge.

<center>Here table 1</center>

It appears that the proximity of "dimerized" $S_2$ bridges leads to a deformation of
the structure, which becomes less stable: the energy raising by 0.21 eV when
"dimerized" $S_2$ bridges are side by side instead of alternated. Besides we found, figure
4a) (S1 to S2), that the activation energy for the separation of a "dimerized" $S_2$ bridge is
0.41 eV, but only 0.18 eV for the reverse process. The first dimerization on the edge is
highly exothermic (-0.76 eV) but there is less stabilization associated with the creation
of the second "dimerized" $S_2$ bridge (-0.23 eV).

<center>Here Fig. 4a) and 4b)</center>

Figures 4a) and 4b) present two different pathways, diverging beyond S2,
respectively with the homolytic dissociation of $H_2$ on the same bridging $S_2$ dimer, or
adjacent dimers.

Figure 4a) represents the computed reaction pathway for the creation of a S
vacancy on the S-edge 100%S. The first step (S1 to S2 via TS1, over a 0.41 eV barrier)

corresponds to the activated "opening" of the S=S bond of one dimer, where the dissociative adsorption of the $H_2$ molecule will further occur following a non concerted mechanism. Then the $H_2$ molecule is slightly physisorbed (-0.02 eV). The total activation energy for the $H_2$ homolytic dissociative adsorption is 0.86 eV but it is lowered if the adsorption occurs directly on a separated bridge (0.65 eV). This is close to the one for the M-edge 50%S (0.48 eV), whereas the former is closer to the one for the M-edge 100%S (0.95 eV) [3].

Figure 4b) points out another pathway for the homolytic dissociative adsorption of $H_2$ on the S-edge 100%S (S2 to S5g). The activation energy (0.92 eV) is of the same order as the first one (0.86 eV), but the chemisorbed state directly accessible after the dissociative adsorption on two neighbouring bridges is far more stable (-1.2 eV) than for the first reaction pathway on a single bridge (-0.48 eV). This value confirms earlier reported results [13, 15, 34] showing that $H_2$ adsorption on this edge is an exothermic process. A panel of configurations is proposed in figure 4b) illustrating H diffusion along the S-edge 100%S. The most stable one (S5b state) occurs when H atoms are bonded on neighbouring $S_2$ bridges alternatively on one side and on the other (-1.25 eV/$H_2$ where the reference is taken as the S1 state). It is of importance to notice that on this edge, for 100% sulfur coverage, hydrogens are not stable on molybdenum atoms, which is not so surprising since Mo atoms are already 6 fold coordinated. Simultaneously, the electronic charge is localized on sulfur as exhibited in table 2.

While one hydrogen bonded on a $S_2$ bridge will decrease its basic character, it can be noticed that the (qualitative) contrast in the acido-basic characters of the bridges in the initial and final configurations gives a trend in the activation energy: when this contrast is strong, e.g. SHSH bridge and $S_2$ bridge (S5a to S5b: Ea = 0.71 eV) or SH bridge and S bridge (S5e to S5d: Ea = 0.65 eV), the reaction is favoured. When the contrast is weaker, the reaction is less favoured, e.g. SSH bridge and $S_2$ bridge (S5g to S5f and reversely: Ea=1.18 and 1.11 eV) or SSH bridge and SSH bridge (S5b to S5a: Ea=1.26 eV). Those diffusion processes are faster than $H_2$ or $H_2S$ associative desorptions, which have almost the same reaction barriers (see figure. 4a)), respectively S3→S2, 1.34 eV, and S4→S7, 1.46 eV)

Here Table 2

The physisorbed molecular state following S2, and precursor to the homolytically dissociated S3 state (see figure 4a)), contains a fraction of the translational and rotational contributions to the free energy prevailing in gas phase. In order to take into account the part of the free energy remaining in this state we apply a fraction of order 0.5, as previously proposed by Dinter et al. for physisorbed $H_2$ and $H_2S$ molecules [3, 42]. (Notice that no physisorbed states of $H_2$ or $H_2S$ were found on M-edge 50%S, see figure 2).

Including the thermodynamics corrections at 575 and 675 K with $pH_2$=10 bar and $pH_2S$=0.1 bar, the S-edge 87%S (S7) is more stable than the hydrogenated S-edge 100%S (S3 to S5a-g) and the non hydrogenated S-edge 100%S (S1 and S2). Rising the

Here Fig. 5

It is worth mentioning that the vibrational contribution to free energy differences between chemisorbed states can reach 0.21 eV (between TS5g and S5b), and as much as 0.57 eV between desorbed and chemisorbed state (between S2 and S5b). It is then clear that omitting this difference would lead to misleading activation free energy barriers and phase diagrams.

*b) initial state S-edge 87% S*

This state is the continuation of the S-edge 100%S after the desorption of one $H_2S$ molecule and relaxation of the edge, so seven sulfur atoms remain on the edge, that is 87.5% sulfur coverage and two "dimerized" $S_2$ bridges (state S10 on figure 6).

We now focus on the effect of one S bridge as first neighbour of the $S_2$ bridge where a $H_2S$ desorption occurs. The left part of figure 6 represents the diffusion of H from a $S_2$ bridge to a S bridge site (S11 to S12) on the S edge 87%S. S12 is the most stable state with H adatoms on the S-edge 87%S. We assume that the barrier to H diffusion inside the S bridge (S12 to S13) is close to that prevailing for the similar process on the S-edge 100%S (S5f to S5c). Finally, the $H_2S$ associative desorption process is carried out between one S bridge site and the vicinal $S_2$ bridge site (S13 to S15).

The S10_bis configuration (not reported here) is similar to S11 without hydrogens. Its energy level is 0.21 eV higher than the S10 configuration, from which it differs by the "opening" of one $S_2$ dimer. This is almost the same difference as on the S edge 100%S between S1 (alternation of "dimerized" bridge and "separated" bridge) and S2 (only one "dimerized" bridge) also previously presented in table 1. It seems that whatever the configuration without H, "dimerizing" every two $S_2$ bridges stabilizes the configurations by 0.2 eV.

On the 87%S edge, hydrogen is more stable on the bridge site between molybdenum atoms supporting the bridging sulfur atom (S12 and S13) than bonded to a sulfur atom of the vicinal $S_2$ bridge, or of the S bridge itself (by about +1 eV from S12, not represented in figure 6). One should notice that as for S-edge 100%S, hydrogen is not stable on a single molybdenum atom (which is 6 fold coordinated).

With thermodynamic corrections included along the reaction path (blue and red lines in figure 6) it appears that the 75%S configuration (S15) is slightly more stable than S10 at 675 K, but this stabilization decreases with decreasing temperature.

Finally, the configuration reached after $H_2S$ desorption is not the most stable of the S-edge 75%S. Indeed, the most stable configuration contains alternating "dimerized" $S_2$ and S bridges (S16) which can be reached through S diffusion on the edge. This latter configuration is slightly more stable ($\Delta G(T) = -0.13$ eV almost independently of temperature) than the former one, and slightly more stable than the S-edge 87%S at 575 K and 675 K.

It is important to notice that the difference in free energy due to the changes in normal vibrational modes can reach 0.32 eV between the chemisorbed states (S14 to S13) and 0.39 eV between chemisorbed and desorbed state (S13 to S15).

The state S16 is S-edge 75%S. In what follows, we will assume, in view of the local similarities of starting configurations (neighbouring $S_2$ dimer and bridging S), that the reduction pathways transforming the two remaining $S_2$ dimers into S bridges, that is from S-edge 75% to S-edge 62%S (Se62S), then from S-edge 62%S to S-edge 50%S (S21) will be similar to that shown on figure 6 to go from S10 to S16. The latter has been therefore detailed because of its prototypical character. Although we have computed the conformations and free energies of the S-edges 75% and 62%, which are included in section 3.2 below, the present assumption will have to be verified when precise barriers and free energy levels of the corresponding intermediates will be needed.

*c) initial state S-edge 50% S*

Figure 7 presents two reaction paths going from $H_2$ homolytic dissociative adsorption (S21 to S22) on S-edge 50%S, to $H_2S$ associative desorption (S23 to S30), and $H_2S$ chemisorption (S30 to S24) to $H_2$ associative desorption (S25 to S21). Notice that the reverse of the latter pathway (S21 to S255) corresponds to heterolytic dissociation of $H_2$ which is then shown, as discussed below, to be preferred over the homolytic pathway on this particular edge. No transition state was found for $H_2S$ desorption, so TS30 and S30 have been displayed with the same energy in the total energy profile.

<p style="text-align:center; color:red;">Here figure 7</p>

The sequence of elementary steps for $H_2$ dissociative adsorption is similar to those presented by Paul *et al.* [7] and Sun *et al.* [8]. The $H_2$ molecule enters the physisorbed S26 state after a 0.2 eV barrier has been passed at the TS26 transition state. Similarly to proposals by Paul and Sun (0.09 eV and 0.23 eV respectively), this step is slightly endothermic by 0.16 eV. Then, the activation barrier for this heterolytic dissociative adsorption is 0.62 eV, pretty close to the value of 0.6 proposed by Paul, and lower than the value of 0.99 eV proposed by Sun. This step is exothermic by -0.19 eV in our case, to be compared to -0.23 eV presented by Paul and 0.02 eV by Sun. The other way, homolytic dissociative adsorption, has not been computed by Paul and Sun, but is unlikely to occur, as there is no physisorbed state, and the activation energy is relatively high compared to the previous one (1.25 eV). The adsorbed state subsequent to this step (S12) is higher in energy by 0.24 eV than S21, therefore homolytic dissociative adsorption is now endothermic.

The sulfur bond energy on the S-edge 50%S is huge (2.09 eV) in comparison to the other S-edges. The sulfur bond energy thus increases with decreasing sulfur coverage as expected by Raybaud *et al.* [31].

Figure 8 presents the reaction path of H diffusion along the S-edge 50%S. H is now more favoured on the bridge site between adjacent molybdenum atoms than bonded to the sulfur atoms (S29a and S29b in comparison to S22). This seems to confirm the basic character of the bridge site lying between 2 molybdenum atoms.

<p style="text-align:center; color:red;">Here figure 8</p>

The free energy differences due to changes in normal vibration modes are still important on this edge. Between chemisorbed states, the difference can reach 0.31 eV (S29b to TS24) and until 0.47 eV between chemisorbed and desorbed states (S29b to S21).

## 3.2    Surface phase diagrams revisited

It is convenient to express the $H_2$ and $H_2S$ pressures (or rather fugacities) in terms of chemical potential. Thus:

$$\Delta\mu_H = \mu_H - \tfrac{1}{2}E_{H2} \text{ and } \Delta\mu_S = \mu_S - \mu_S(bulk) \qquad [7]$$

with $\mu_H$ and $\mu_S$ computed according to the method detailed in section 2.2, and $E_{H2}$ and $\mu_S(bulk)$, respectively the total energy of $H_2$ computed in DFT and the total energy of sulfur in alpha crystalline phase.

In what follows, "relevant HDS conditions" are defined in terms of temperature interval (573-700K), partial pressure of $H_2$ p($H_2$) (1-200 bar), and ratio of partial pressures p($H_2S$)/p($H_2$) ($10^{-4}$-$10^{-1}$). Indeed, industrial hydrotreating units are operated at

p($H_2$) ranging from a few bar (HDS of naphtas as pretreatment of reforming feedstocks) to almost 200 bar (hydroconversion of vacuum residua), and temperature ranging from about 573K to about 700K. The ratio p($H_2S$)/p($H_2$) lies most often between 0.01 and 0.05. Besides, experiment performed on model molecules at the laboratory scale, or spectroscopic experiments may include p($H_2$) slightly over one atmosphere (taking into account pressure drops across small fixed beds of catalysts), and in general p($H_2S$) might be very low, however we assume they will stay above the limit of stability of $MoS_2$ with respect to Mo metal. Finally, our definition of "relevant HDS conditions" tries to encompass all these situations corresponding to a possible experiment involving a $MoS_2$ based catalyst.

In figure 9 are plotted the grand potentials of several M-edge configurations relative to the grand potential of the S1 state, against the chemical potential of sulfur relative to its alpha cristalline phase, $\Delta\mu_S = \mu_S - E_S^{bulk}$. Here, these grand potentials correspond to the free enthalpies of reactions connecting S1 to the other states, $\Delta G(T, p(H_2), p(H_2S))$, plotted for T=623K and p($H_2$)=10 bars. A large set of configurations was simulated for each type of edge (S coverages and H coverages), but for clarity only the most stable for each type of edge are reported. This figure displays the range of stability of non-hydrogenated M-edge 50%S states, and the M-edge 50%S+50%H with 2 H on the surface (S4_bis state, see figure 3).

One can notice that at low chemical potential of sulfur, below -1.22 eV (i.e. $H_2S$ pressure <560 Pa), the M-edge 37%S (S6) becomes the most stable. However, its hydrogenated counterpart, Me37S_Hb_S_SH, is not stable. This is partly due to the loss of entropy in the chemisorbed state of H: the vibration free energy is higher for the hydrogenated edge state than for the other one, however it is still compensated by the loss in translation and rotation entropies. This means that the probability that hydrogen remains adsorbed at the sulfur-vacancy site is very small, so that this site shall remain available for adsorption of other reactants.

Here Figure 9

Here figure 11

The phase diagram at 623 K shown on figure 11, reveals the stability of 3 configurations in relevant HDS conditions, the S1 state, the S4_bis state, and the S6 state, consistently with figure 9. At very high $p(H_2S)/p(H_2)$ ratio ($\Delta\mu_S > -0.33$ eV) the M-edge 62%S becomes the more stable, but this is far out of relevant HDS conditions. The stability of a sulfur vacancy on the M-edge (S6) is consistent with previous DFT calculations [36, 37] showing that a coordinately unsaturated Mo site (Mo-CUS) can be stabilized in relevant HDS conditions. This corresponds to a S coverage on the surface around 37%. Lower S coverages explored so far, M-edge 25%S in the present work, or M-edge 33%S [31], are not stabilized, which means that a good estimate of S coverage at the edge is close to 37%S corresponding to 50% Mo-CUS.

At 623 K, H is thermodynamically stable on the M-edge above 5.3 bar of $H_2$ pressure, and the configuration with the anion vacancy (S6) is stable for a ratio $p(H_2S)/p(H_2) < 10^{-3}$. A decrease of temperature (473 K in figure 13 a)) extends the stability domain of the S4_bis state towards lower values of $H_2$ pressure (now stable above $p(H_2)=0,2$ bar) and lessens the domain of the S6 state, (now stable below $p(H_2S)/p(H_2)\approx2.10^{-5}$).

The coexistence of the S4_bis and S6 configurations is expected to favour HDS reactions, since the S6 state presents an adsorption site for sulphided molecules, which could then undergo hydrogenation and hydrogenolysis thanks to available H adatoms diffusing from the S4_bis state.

The two symbols in figure 11 represent the following conditions: temperature = 623 K and p(H$_2$S) = 0.1 bar, and p(H$_2$) = 10 bars for the crossed full square, and p(H$_2$) = 100 bars for the full triangle. These conditions correspond to the purple dashes for the S1 and the S6 states in the energy profile, figure 2. The two figures are consistent since the "M-edge" 37%S (S6 state) becomes more stable than the M-edge 50%S (S1 state) when increasing the H$_2$ pressure above about 100 bars.

<p style="text-align:center;color:red;">Here Figure 10</p>

<p style="text-align:center;color:red;">Here Figure 12</p>

In figure 10 are plotted the Gibbs energies of all the most stable computed configurations for each kind of S-edge (S coverage, H coverage) against the sulfur chemical potential $\Delta\mu_S$, at 623 K and p(H$_2$)=10bar. We observe that S-edge 100%S+50%H with 4 hydrogens on the edge (Se100S Sh_SH_sh_SH is the most stable for high $\Delta\mu_S$ (high H$_2$S pressure), whereas S-edge 50%S (Se50S or S21) is the most stable for low $\Delta\mu_S$ (low H$_2$S pressure), and S-edge 62%S (Se62S, see figure 5c)) is the most stable in between.

Figure 12a) depicts a phase diagram of the S-edge at 623 K. In relevant HDS conditions (p(H$_2$) between 1 bar and 200 bar, and p(H$_2$S)/p(H$_2$)< 0.1), that is inside the domain delimited by dashed lines, all the S-edge configurations represented in the diagram may be reached except 100%S. Changes in temperature, H$_2$ pressure, or H$_2$S pressure, will change the stable phase, and these changes will change the sulfur coverage on the edge. Each S bridge is an anion vacancy site potentially able to accept another sulfur to form a $S_2$ bridge. S-edge coverage changes from one stable state to another upon a change of reaction conditions.

The phase diagram in figure 12a) exhibits a stability domain for the S-edge 100%S until a maximal $H_2$ pressure of 0.8 bar ($\Delta\mu_H$=-0.41 eV) and a minimal $p(H_2S)/p(H_2)$ ratio of 0.4 ($\Delta\mu_S$=-0.88 eV). Then if the $H_2$ pressure is further increased, H chemisorption becomes strongly exothermic, and for dissociative adsorption of up to two $H_2$ molecules on one S-edge, the more H chemisorbed on the edge, the more stable the edge. Then at a constant $H_2$ pressure, decreasing the $H_2S$ pressure favors the $H_2S$ desorption from the S-edge 100%S 50%H, to S-edge 62%S or even S-edge 50%S. The state S-edge 62%S is stable down to $p(H_2S)/p(H_2)$=0.02 ($\Delta\mu_S$= -1.03 eV). All these states have domains of stability intersecting the domain of relevant HDS conditions.

The crossed square dots in figure 11 and 12 represent a particular HDS condition, i.e. 623K, $p(H_2)$=10 bar and $p(H_2S)/p(H_2)$=0.01. This particular HDS environment stabilizes the S-edge 50%S and the M-edge 50%S+50%H (see figure 11). In these conditions hydrogen is stable solely on the M-edge.

Here Figure 13

In figure 13 are reported the phase diagrams for the M-edge and S-edge in HDS conditions at 473 and 573 K. The comparison with phase diagrams at 623 K of figures 11 and 12 points out the monotonic behaviour of the evolution of the stability domains for the different surface phases with temperature.

On the M-edge, an increase of temperature tends to reduce the stability of chemisorbed hydrogen in relevant HDS conditions. The stability of a sulphur vacancy on the edge (S6) increases with temperature, and at 623 K this domain is extended until a maximal $pH_2S/pH_2$ ratio of $10^{-3}$, and a $H_2$ pressure up to 40 bars at $pH_2S/pH_2$ $10^{-4}$ .

On the S-edge, at the lower temperature (473K), the 100%S+50%H edge holds on a large domain of pressure, in the relevant HDS conditions (inside the area delimited by dashed lines). It remains stable even for a $H_2$ pressure lower than 0.1 bar. The rest of the domain of relevant HDS conditions is filled by the S-edge 50%S. Notice that the S-edge 100%S is stabilized by the presence of hydrogen on the edge. When increasing the temperature, this domain and the 62%S domain are pushed towards larger values of the $pH_2S/pH_2$ ratio. This evolution (figure 12a), 13b), 13d)) is observed through the change of the different domains containing the crossed square dot ($pH_2$=10 bar, $pH_2S$=0.1 bar).

**3.3**    ***Normal modes of vibration for edge SH and MoH groups***

As indicated above, the vibration frequencies of normal modes were computed in order to calculate the phase diagrams. As a consistency test, it is worth comparing predicted and experimental frequencies, since the latter should reflect the presence of species at the conditions of preparation of the studied samples for spectroscopic experiments.

This comparison is presented in Table 3. The 529 cm$^{-1}$ vibration frequency of the S-S dimer on the S-edge has been brought out by Raman spectroscopy by Polz et al. [38] in experimental conditions for which we predict the coexistence of S-edge 62%S and M-edge 50%%S (S1) (573K, P(H$_2$) 350-400 mbar, $\Delta\mu_S$ = -0.89 eV). This frequency is in close agreement with our calculation (536 cm$^{-1}$) for the dimer belonging to the S-edge 100%S 25%H (S5g). Besides its presence on the S-edge 100%S, this S-S dimer also exists on the S-edge 62% and its computed vibration frequency now corresponds to 548 cm$^{-1}$, also very close to the experimental one. The agreement between the theoretical and the experimental frequencies for these particular conditions (575K, $\Delta\mu_S$=-0,89 eV, pH$_2$=300 mbar) is therefore consistent with the predicted stability of the S-edge 62%S in our phase diagram.

Here table 3

Our computation of the normal modes for several configurations shows that, consistently with experimental data [39] and corroborated by other theoretical works [40, 14], the S-H bond stretching vibration frequencies are around 2500 cm$^{-1}$ both on S-edge and M-edge. Inelastic Neutron Scattering (INS) spectroscopy undertaken by Sundberg et al. [39] enables to assess indirectly vibration frequencies of edges at 473K and under 1 bar, 20 bar or 50 bar of H$_2$ pressure. After pre-sulfidation at 573 K with a pH$_2$S/pH$_2$ ratio of 10 % and two hours in the conditions mentioned previously, first the temperature was decreased to 273 K, then the chamber was evacuated for one hour to remove H$_2$ from the gas phase, and finally the sample was cooled down to 60K before beginning the spectroscopy measurements. The bending vibration frequency of S-H bonds reported at 650 cm$^{-1}$ with INS spectroscopy corresponds in our table to the frequencies between 490 and 690 cm$^{-1}$. The theoretical bending frequencies of the S-H bond on the S-edge 100%S+50%H (676-658 cm-1) is in relatively good agreement with the experimental data since we find that the thermodynamically stable domain of the S-edge 100%S+50%H extends towards the lower values of the ratio pH$_2$S/pH$_2$ with the decrease of temperature. On the M-edge the frequencies of the bending vibration of the S-H bond found on the most stable hydogenated edge (S4_bis: 644-496 cm$^{-1}$) are somewhat lower than the typical experimental frequency. This discrepancy could be due to the harmonic model, or to the finite difference step size (in our case 0.02 Å) used to compute the Hessian Matrix. The uncertainty on this last parameter for our system was estimated at most to be 20 cm$^{-1}$ (step size between 0.01 Å and 0.02 Å) for real frequencies, so it should not be solely responsible for the discrepancy. The interaction between hydrogens for the S4_bis state affects the vibration frequencies, especially for the vibration along the axis formed by the two hydrogens, which corresponds to the lowest S-H bending frequency (496 and 505 cm$^{-1}$). On the M-edge 50%S the diffusion barriers for surface hydrogen atoms are low, so it is expected that H diffusion occurs on the edge surface. Then for larger distances between H ad-atoms, and therefore reduced interaction, the predicted vibration will be closer to the experimental value. Whatever the bending vibration frequencies, the experimental data bring out the stability of hydrogen on the edges in the following conditions T=473K, p(H$_2$)= 1, 20, or 50 bar, and a 10% p(H$_2$S)/p(H$_2$) ratio, which is in agreement with our phase diagram figure 13 b) (S-edge 100%S+50%H) and 13 a) (M-edge 50%S+50%H, S4_bis) for both edges.

Moreover we have identified, e.g. for state S2, vibrations of Mo-H groups ranging between 1800 cm$^{-1}$ and 1900 cm$^{-1}$. A similar mode was only observed and assigned for metal hydride/sulfur complexes, by Burrow et al. [41], but was never observed, as far as we know, on MoS$_2$-based catalysts. Since S2 is never a predicted stable state in usual conditions of observation, this lack of occurrence supports the higher stability of the S-H bond compared to the Mo-H bond, as predicted from our phase diagrams. The other situation for which we find a Mo-H bond is S25 (S-edge 50%S+25%H), which is not a stable state according to our prediction. For the stretching mode of this bond, we predict frequencies 1294 cm$^{-1}$ and 1370 cm$^{-1}$. These frequencies have never been observed experimentally, in agreement with our stability prediction.

From this analysis, we conclude that our computational predictions are in good agreement with the vibrational spectroscopy data released so far. We did not find contradictions from this rather stringent test of theoretical results.

## 3.4 Bronsted-Evans-Polanyi relationships (BEPR)

Since a lot of activation barriers for adsorption and desorption processes have been brought out along this study, it is interesting to explore if Bronsted-Evans-Polanyi relationships (BEPR) may exist. It is worth to mention that when a BEPR exists for a direct reaction step, another BEPR can be derived for the reverse reaction. Given $Ea_d$ the activation energy for the direct process and $\Delta E_d$ the corresponding reaction energy, the BEPR states that:

$$Ea_d = \lambda E_d + \beta \qquad [8]$$

For the reverse reaction, due to energy conservation, $\Delta E_r = -\Delta E_d$, and $Ea_r = Ea_d + \Delta E_r = Ea_d - \Delta E_d$, therefore one obtains:

$$Ea_r = (1 - \lambda)\Delta E_r + \beta \qquad [9]$$

So knowing the equation for one reaction, we can easily find the equation for the reverse reaction.

As presented in Fig. 14, relatively good correlations are obtained for $H_2$ and $H_2S$ associative desorptions, supporting the existence of BEPR for such processes and their reverses.

We should mention that for diffusion processes, despite all the barriers calculated along reaction pathways, no clear BEPR was found. This is due to the dependency of the activation energy on several parameters during diffusion processes. In particular, the bond strength and the bond stretching before creation of another bond are responsible for the location of the transition state (TS) along the reaction pathway (determining an early or late transition state [35]). Besides, the edge structure relaxation, and the repulsive interaction between hydrogens will affect the activation energy.

<center>Here Figure 14</center>

### 3.4.1   *Associative desorption of $H_2$*

For such processes, the relation between activation energy and reaction energy is well described by the BEPR straight line except for 2 outliers at $\Delta E$ = -0.24 eV and $\Delta E$ = 0.12 eV. These 2 points correspond respectively to the reverses of a homolytic dissociation on S-edge 50%S and a heterolytic dissociation on M-edge 50%S. It has been shown in the recent work of Van Santen *et al.* [35] that the BEPR depends on the site where the reaction occurs.

In our case the associative desorption reaction on the M-edge related to one of these outliers involves one molybdenum and one sulfur (heterosynthetic: S2 to S1 via TS1, see figure 2) unlike other associative desorptions on the M-edge, where desorption involves two sulphurs (homosynthetic). Lower stretching frequencies of Mo-H bonds with respect to S-H bonds (see Table 3) already reflect a lower (restoring) force constant in the first case (weaker local curvature of the potential energy surface), while the opposite partial charges born by hydrogens (Mo-H$^{\delta-}$ and S-H$^{\delta+}$) are also in favour of the observed lower activation energy for heterosynthetic association at comparable energy of reaction. .

Concerning the homosynthetic associative desorption of $H_2$ from the S-edge 50%S, the main difference with other homosynthetic associations on the S-edge is the lower coordination of Mo atoms, which tends to localize more electronic density on terminal S atoms. According to the Bader analysis [26, 27] of the electronic density of the atoms on the hydrogen-free edges, Table 2, the total charge on the S atom is -0.96e for the S-edge 50%S (S21 in figure 7) whereas on the S-edge 100%S (S2 in figure 4a), the total charge on the separated sulfur atoms is -0.78e. At the transition states (TS21 for S-edge 50%S, figure 7, and TS2 for S-edge 100%S, figure 4a), the total charges on the sulfur atoms are -0.76e and -0.62e for S-edge 50%S and S-edge 100%S respectively. During the process of associative desorption, the dominantly ionic S-H bond is stretched, but the attractive (mostly electrostatic) interaction between hydrogen and sulfur is larger for S-edge 50%S since more charge resides on sulfur. Hereby the bonding between sulfur and hydrogen is less extended, resulting in a TS state with a shorter S-H bond (1.58 Å on the S-edge 50%S versus 1.79 Å on the S-edge 100%S). This effect is responsible for a larger H-H distance at the TS state (1.11 Å for the S-edge 50%S versus 0.92 Å for the S-edge 100%S). This difference in the H-H distance reflects a 0.75 eV increase in the binding energy of $H_2$ at the TS on S-edge 100%S versus S-edge 50%S. Accordingly, the activation of $H_2$ homosynthetic associative desorption is less favoured on S-edge 50%S than predicted by the BEPR which holds for homosynthetic associative desorption on S-edge 100%.

In the case of the heterosynthetic associative desorption from molybdenum and sulfur on the S-edge 50%S (S25 to S26 via TS25, figure 8, $\Delta E=0.35$ eV, barrier 0.97 eV) the two effects counteract (more attraction of $H^{\delta+}$ by $S^{\delta-}$, less attraction of $H^{\delta-}$ by $Mo^{\delta+}$), and the BEPR main tendency is followed again.

### 3.4.2   Desorption of $H_2S$

It is worth to notice that in order to check if a BEPR exists, it is necessary to take into account the energy of elementary processes, as stated by Van Santen *et al.* [35] as well as the energies of transition state and pre-transition state. The change of activation energy for desorption of $H_2S$ as function of the reaction energy is well described by a BEPR very close to the main diagonal ( $\lambda = 1, \beta = 0$ ) except for one value which corresponds to the homosynthetic associative desorption of $H_2S$ from the M-edge 100%S (data from [3]). The pre-transition state corresponds then to two SH groups bonded to the same Mo and interacting through a H...SH hydrogen bond. For the other $H_2S$ desorption processes considered either on the M-edge 50% or S-edges, the pre-transition state is always molecularly chemisorbed $H_2S$, bonded to one (top) or two (bridging) Mo ions. The associative process precursor to molecularly chemisorbed $H_2S$ is now the limiting step for the overall associative desorption, in particular for a process with a low desorption energy as S6 to S7 (figure 4a)) on the S-edge 100%S, and S14 to S15 (figure 6) on the S-edge 87%S. .

One would have expected that the desorption energy should be inversely proportional to the S coverage, since as mentioned in section 3.1.2.1.c, the S bond energy is inversely proportional to the S coverage. But since the associative desorption may be decomposed into an associative pre-process and a desorption process, the latter is associated to the former, and there is no rule of proportionality between S coverage and the $H_2S$ desorption energy itself.

Considering all points including outliers, The BEPR straight lines for $H_2$ and $H_2S$ desorption are rather parallel (slopes are almost the same 0.79 and 0.83), however the activation energy for $H_2$ desorption is always higher than the activation energy for $H_2S$ desorption. This means that, if we consider dissociative adsorption and associative desorption on the same site, when the binding energy of sulfur on the edge is lower than ~0.8 eV (0.88/0.79-0.22/0.83), then the desorption of $H_2S$ is faster than that of $H_2$.

# 4    Discussion

## 4.1    Comparison with previously proposed surface phase diagrams for $MoS_2$ catalytic edges

The phase diagram at 623 K for the M-edge presented in figure 11 modifies the stability domains of hydrogen previously determined by Cristol *et al.* [13]. In agreement with Bollinger *et al.* and Lauritsen *et al.* [15, 16], the stability of H on the edges is expected to be extended to less reducing conditions (P($H_2$)=5 bars corresponding to $\Delta\mu_H$=-0.35 eV). Then in contrast to Cristol and Bollinger, the stability domain of the Mo-edge at 50% sulfur coverage is surrounded by Mo-edge 50%S+50%H and Mo-edge 37% S for $\frac{p(H2S)}{p(H2)}=10^{-3}$ (corresponding to $\Delta\mu_S$=-1.2 eV). Those states are stable in relevant HDS conditions since they are overlapped by the domain of relevant HDS conditions delimited by the dashed lines.

At this point it is important to underline the significance of including thermal and entropic corrections in surface phase diagrams: the comparison of figures 11a) and 11b), and 12a) and 12b), is illustrative, showing that in relevant HDS conditions, the correct stable states may be missed out without such corrections.

## 4.2 Significance of our results with respect to the activation of dihydrogen by MoS$_2$ based HDS catalysts

In section 3.1 were reported the activation energies for the heterolytic dissociative adsorption of H$_2$ on the M-edge 50%S (0.48 eV) and on the S-edge 50%S (0.62 eV), and the activation energies for the homolytic adsorption on the M-edge 100%S (0.95 eV) [3], S-edge 50%S (1.25 eV), and S-edge 100%S (0.65 eV). The phase diagram reported by Schweiger *et al*. [36] showed that S-edge and Mo-edge are in competition to determine the morphology of catalytic nanoparticles. The lowest activation energy is the one for the M-edge 50%S and it has been shown on figure 11 that the resulting hydrogenated edge (M-edge 50%S+50%H) may coexist with the 37%S hydrogen-free M-edge. Knowing the relative free energies of these states, it is possible to evaluate the fraction of sulfur vacancies at bridge sites of the M-edge 50%S+50%H (S4-bis) (or fraction of M-edge 37%S). Starting from 100 % of lacunar M-edge bridge sites at 575 K under 10 bar H$_2$ below 10$^{-3}$ bar H$_2$S, this coverage decreases to 0.15 % under 0.1 bar H$_2$S, but increases again to circa 2% if temperature is then raised to 675K. The M-edge 37%S should favour the chemisorption at the anionic vacancies, of the sulfur adatom from thiophenic derivatives present in crude oil fractions. Subsequently, hydrogens adatoms on the same edge are expected to be responsible for the hydrogenation and hydrodesulfuration of chemisorbed thiophenic derivatives.

On the S-edge 50%S, $H_2$ heterolytic dissociative adsorption subsequent to physisorption is exothermic but endergonic. On this same edge, $H_2S$ associative desorption is the slowest process compared to that occurring on all other edges. Nevertheless this does not mean that this edge does not play a role during HDS, since it is stable in relevant HDS conditions (see figure 12a), and 13a), 13c)) and sulfur adsorption should occur, as the activation energy for $H_2S$ adsorption is expected to be around 0.1 eV. This value is calculated from the BEPR established in section 3.4 with the energy difference:

$$\Delta E = (E_{Me50\%S} + E_{H2S}) - E_{62\%S+H}$$

where $E_{62\%S+H}$ is the hydrogenated S-edge 62%S DFT energy, $E_{Me50\%S}$ is the DFT energy of the S-edge 50%S free of H, and $E_{H2S}$ the DFT energy of $H_2S$ in gas phase. This edge could also enable the adsorption of a thiophenic derivative if the steric effects are not too large.

The emerging picture for the lower ratios $p(H_2S)/p(H_2)$ inside the range of relevant HDS conditions (left part of the dashed rectangles in figures 11 to 13) is then a situation where activated hydrogen is provided by the M-edge, while anionic vacancies are abundant on the S-edge, and scarce on the M-edge. The chemisorption of thiophenic compounds should be possible on these vacancies, but in strong competition with $H_2S$,

For higher ratios $p(H_2S)/p(H_2)$ inside the range of relevant HDS conditions (right part of the dashed rectangles in figures 11 to 13), and for instance with reference to the conditions 10 bar of $H_2$ pressure and p($H_2$S)/($H_2$) ratio of 0.1 brought forward by Bollinger *et al.* [15], [16], and represented by the crossed square dots in figures 11 to 13, the situation changes somewhat. The stable edge surfaces of $MoS_2$ , upon spanning the relevant range of from lower to higher operating temperatures (475K to 700K), are M-edge 50%S+50%H (S4_bis) together with S-edge 100%S+50%H at 475K, then M-edge 50%S+50%H (S4_bis) together with S-edge 62%S at 575K, then M-edge 50%S+50%H (S4_bis) together with S-edge 50%S at 623K. In other terms, $H_2$ is dissociatively activated on both edges at the lower temperature, and as temperature increases, it remains available on the M-edge only, while being depleted from the S-edge, which also looses sulfur, offering more and more anionic vacancies. Still for these conditions, but above 648K , the stable M-edge will become S1 (M-edge 50%S) also depleted in H : it then questionable wether $MoS_2$ will retain any catalytic activity in hydrogenation and/or hydrogenolysis. The stability of M-edge 50%S + 50%H will be of course recovered, and thereefore the activity, likely, if p($H_2$) is increased beyond some threshold, the higher the higher the operating temperature.

**5   Conclusions**

In this work, we have investigated a set of various reaction pathways for $H_2$S and $H_2$ dissociative adsorption and associative desorption, as well as various hydrogen diffusion processes occuring at both sulfur and molybdenum edges of $MoS_2$, in conditions encompassing those of industrial hydrotreating  .

For the first time in this context, DFT computations have been carried out for this system including thermal and entropic effects of the gas phase and catalytic edges. It has been shown that these effects have a significant impact on the activation free energies and on stabilities of intermediate states, hereby on and surface phase diagrams and transition kinetics .

In relevant HDS conditions, thermodynamically stable chemisorbed H states have been brought out on the M-edge 50%S and the S-edge 100%S. On the latter, for instance, at 623K, the hydrogenated edge is stable for $p(H_2S)/p(H_2)$ ratios above 0.01. On the former, for instance, the hydrogenated edge is stable beyond 5 bar of $H_2$ pressure and $p(H_2S)/p(H_2) = 0.001$. On the latter, at 623K and $p(H_2) = 5$ bar, the hydrogenated edge becomes stable only above $p(H_2S)/p(H_2) = 0.03$. Moreover it has been shown that the domains of stability of these hydrogenated edges extends to lower $H_2$ pressures and lower $p(H_2S)/p(H_2)$ ratios when temperature decreases.

The possible coexistence of the hydrogenated M-edge 50%S (Me50%S+50%H) and of the M-edge 50%S containing an anion vacancy (Me37S) has been evidenced. .When for instance 100 % of M-edge bridge sites are lacunar at 575 K under 10 bar $H_2$ below $10^{-3}$ bar $H_2S$, this coverage decreases to 0.15 % under 0.1 bar $H_2S$, or circa 2% at 675K. These sulfur vacancies, although scarce on the M-edge, are abundant on the S-edge, and are expected to play an important role in the catalysis of hydrodesulfuration reactions in presence of un-promoted $MoS_2$

Since the hydrogenated S-edges 50%S+(50-100%H)  are not thermodynamically stable according to the phase diagram, it should be interesting to verify to which extent they can show up as metastable states.

A comparison between experimental measurements reported in the literature, and our theoretical calculations of vibration frequencies allowed to identify S-H stretching and bending, and S-S stretching modes, respectively at 2500, 650 and 530 cm$^{-1}$, belonging to surface species at both edges. The calculated vibration frequencies corresponding to the theoretically predicted stable surface species at conditions of measurements for the available spectroscopic experiments in literature are in agreement with observed frequencies. .

Several activation energies for $H_2$ dissociative adsorption have been computed for the main edges present in the relevant HDS conditions, i.e. S-edge 100%S, S-edge 50%S, and M-edge 50%S. The lowest barrier has been found for the $H_2$ dissociative adsorption process on the M-edge 50%S (0.48 eV). On the other edges the barriers are slightly higher, with 0.65 or 0.86 eV and 0.78 eV respectively for S-edge 100% S starting from "separated" or "dimerized" $S_2$ bridge and S-edge 50%S.

The preferred pathways for dissociative association of $H_2$ are heterolytic on the M-edge 50%S, and S-edge 50%S, and homolytic on the S-edge 100%S, 87% S and generally as long as $S_2$ dimers are attacked.

The activation energies for $H_2S$ associative desorption have been also computed for S-edge 100%S, S-edge50%S and M-edge50%S, respectively at 2.07, 2.51 and 1.92 eV. Vibrational entropic corrections up to 0.6 eV depending on temperature, have to be taken into account for estimates of free energy barriers.

Two Brönsted-Evans-Polanyi Relationships have been determined for the associative desorption of $H_2$ and $H_2S$, and outliers to these BEPRs have been discussed.

We think that this work provides a well quantified determination of the chemical active species at the edges of the $MoS_2$ nano-crystallites in HDS conditions. A great attention was paid to provide the stability domains of CUS, sulfhydryl group, sulfur anions as a function of temperature and partial pressures of $H_2S$, $H_2$. Moreover the kinetic properties of these species were also evaluated: hydrogen dissociative adsorption, hydrogen diffusion, and CUS creation associated to $H_2S$ desorption. We hope that this will help for a more accurate investigation of the elementary steps of HDS mechanisms where the same active species are involved.

Finally the question remains open of the influence of kinetics on the edges meta-stabilities. However, in view of the large number of interacting processes involved, and in order to gather statistically meaningful data over large enough time lags and site numbers, we expect first-principles based kinetic Monte Carlo simulations to be an efficient approach to address these questions. Considering such perspectives, we hope that the present work provides an appropriately extensive database of DFT barriers, complementing consistently our previous contributions ([3, 42].

**ACKNOWLEDGMENTS**

This work has been performed within the SIRE project (Grant No. ANR-06-CIS6-014-04) sponsored by the Agence Nationale de la Recherche (ANR).

**List of tables**

**Table 1:** Energy differences for several configurations on the S-edge 100%S

**Table 2:** Bader analysis of the electronic charge born by atoms contained in the $MoS_2$ catalyst sheets.

**Table 3:** Comparison of the experimental and theoretical normal modes. [(1) corresponds to Inelastic Neutron Scattering (INS) experiments by [39], (2) to Raman spectroscopy experiments by [38], and (3) to Infra-Red spectroscopy experiments by [41]].

| configurations | Energy [eV] | | models |
|---|---|---|---|
| | our work | Reference (*) | |
| no "dimerized" S2 bridge | 0.00 | 0.00 | 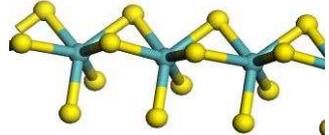 |
| one "dimerized" S2 | -0.76 | | 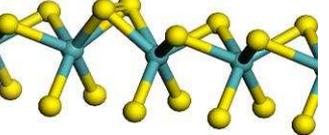 |
| 2 "dimerized" S2 bridge alternately | -0.99 | -1.16 | 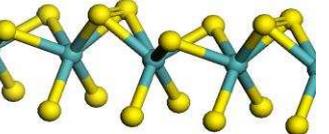 |
| side by side | -0.78 | | 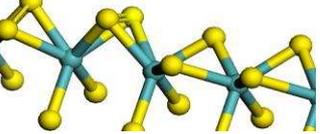 |
| no "dimerized" S2 bridge | -0.67 | -1.06 | 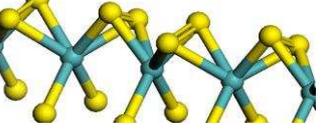 |

**Table 1**: Energy differences for different configurations of the S-edge 100%S, where the edge without dimer is taken as a reference for the energy. (*) corresponds to the reference [34] where the cell contains 2 units of $MoS_2$ on the edge instead of 4 in our case.

|  | Bader analyis of the valence charge | Net Charge on atoms |
|---|---|---|
| S-edge 100%S | | |
| • S at the edge: | | |
| Dimerized bridge | 6.40 e | -0.40 e |
| Separated bridge | 6.78 e | -0.78 e |
| S in the inert plane | 6.85 e | -0.85 e |
| • Mo near the edge | 4.21 e | +1.79 e |
| Mo in the inert plane | 4.30 e | +1.70 e |
| S-edge 50%S | | |
| • S at the edge: | 6.96 e | -0.96 e |
| S in the inert plane | 6.85 e | -0.85 e |
| • Mo near the edge | 4.21 e | +1.79 e |
| Mo in the inert plane | 4.30 e | +1.70 e |

**Table 2**: Bader analysis of the electronic charge borne by Mo and S atoms contained in the $MoS_2$ catalyst sheets. Since the pseudopotential we used contained 6 valence electrons for each atoms, the net charge on each atom is deduced from the electron valence charge of the Bader analysis [26, 27].

| reference | S-H | S-S | Mo-H | Experimental conditions/Model |
|---|---|---|---|---|
| Sundberg *et al.* (1) INS | 2500(s) 650(b) | | | T=573 under p($H_2$S)/p($H_2$)=10%, then T=473 under H2 gas (1, 20, 50 bar) |
| Polz *et al. (2)* Raman | | 529(s) | | T=573K ; p($H_2$)=300-450mbar; p($H_2$S)/p($H_2$)=1/9 <=>$\Delta\mu_s$= -0,89eV |
| Burrow *et al.* (3) IR spectroscopy | | | 1740(s) 1858(s) | Mo(H)(tipt)$_3$(PMe$_2$Ph)$_2$ Mo(H)(tipt)$_3$(PEtPh$_2$)    p=1bar , T=293 K |
| This work Se100%S | 2598,2591(s) 685 ,664 (b) 655, 653 (b) | 536 (s) | | 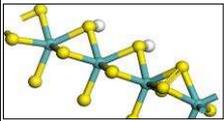 S5g |
| | 2520,2517(s) 676-658(b) | | | 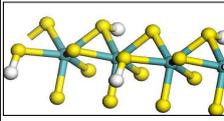 100%S+50%H |
| Se50%S | 2582,2524(s) 631, 574 (b) 542, 442 (b) | | | 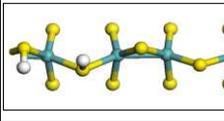 S22 |
| | 2531 (s) | | 1374,1285(s) | 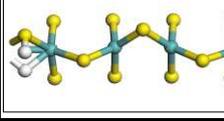 S25 |
| Me50%S | 2587 (s) 662, 519 (b) | | 1871 (s) 736, 617 (b) | 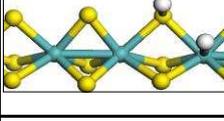 S2 |
| | 2583,2570(s) 639, 561 (b) 532, 489 (b) | | | 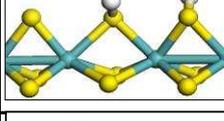 S4 |
| | 2573,2567(s) 644, 640 (b) 505, 496 (b) | | | 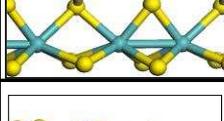 S4_bis |
| Me100%S | 2590,2586(s) 624, 609 (b) 511, 506 (b) | 576 (s) | | 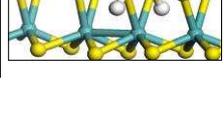 100%S+25%H |

**Table 3**: Comparison of the experimental and theoretical normal modes. (1) corresponds to Inelastic Neutron Scattering (INS) experiments by [39], (2) to Raman spectroscopy experiments by [38], and (3) to Infra-Red spectroscopy experiments by [41]].

**List of figures**

**Figure 1:** A typical periodic slab model for $MoS_2$ S-edge, showing coverage by four $S_2$ bridges. Legend: (color on line) black spheres (turquoise): molybdenum atoms, grey spheres (yellow): sulfur atoms.

**Figure 2**: Reaction pathway for the creation of one S vacancy on the M-edge 50%S. Blue line: Free energy at 575K, $pH_2$=10 bar, $pH_2S$=0.1 bar ($\Delta\mu_S$=-1.01 eV). Red line: Free energy at 675K, $pH_2$=10 bar, $pH_2S$=0.1 bar ($\Delta\mu_S$=-1.13 eV). Energies of the S1 state are taken as references for each reaction path. On the ordinates axis is reported (purple dash) at 623 K the free energy of the S1 state at 100 bars and 10 bars of $H_2$ pressure and the S6 at 0.1 bar of $H_2S$ pressure ($pH_2$=10bar). Also the total energy, the free energy at 575 K and the free energy at 675 K of the S4_bis state (see figure 3) are reported.

**Figure 3**: Configuration of the most stable state on the M-edge 50%S (-1.09 eV), the S4_bis state, also called Me50S_SH_S_SH.

**Figure 4 a):** Reaction pathway for the $H_2S$ desorption process on the S edge 100%S, through $H_2$ adsorption on one bridge. The first step is the separation of the 'dimerized' Sulfur in the bridge. Blue line: Free energy at 575K, $pH_2$=10 bar, $pH_2S$=0.1 bar. Red line: Free energy at 675K, $pH_2$=10 bar, $pH_2S$=0.1 bar. Energies of the S1 state are taken as references for each reaction path. **b):** $H_2$ adsorption on two neighboring bridges and diffusion along the S edge 100%S (S2 and S5a states are equivalent to those on figure 4a). Blue line: Free energy at 575K, $pH_2$=10 bar, $pH_2S$=0.1 bar. Red line: Free energy at 675K, $pH_2$=10 bar, $pH_2S$=0.1 bar. Energies of the S1 state (in figure 4a)) are taken as references for each reaction path. Note that the most stable state at 0K is S5b, while at 675K (red path) it becomes S7.

**Figure 5**: a) S-edge 100%S+100%H and b) S-edge 100%S+50%H, respectively 1.3 eV and 2.6 eV lower in energy than S1 state with $H_2$ in gas phase. i.e. -0.325 eV and -1.3 eV adsorption energy per $H_2$ molecule. c) configuration of the S-edge 62%S involved in the phase diagram of figure 11.

**Figure 6:** Reaction path for $H_2S$ desorption from the 87%S-edge. Blue line: Free energy at 575K, $pH_2$=10 bar, $pH_2S$=0.1 bar. Red line: Free energy at 675K, $pH_2$=10 bar, $pH_2S$=0.1 bar. Each reference energy is the energy of the S10 state for each reaction path. Note that the S15 configuration is less favourable by $\Delta G(T) = $ -0.13 eV (the dependence in temperature in the range of interest is less than 0.01 eV) than S16, where 'dimerized' $S_2$ bridge alternate with S bridges on the edge. (For simplification, $H_2$ molecule in inset S10 and $H_2S$ molecule in insets S16 have been omitted)

**Figure 7:** Reaction path on the S-edge 50%S for the $H_2$ homolytic dissociative adsorption on two neighboring S bridges (S21 to S22), then $H_2S$ desorption (S23 to S30), followed by $H_2S$ adsorption (S30 to S24), eventually $H_2$ heterolytic associative desorption via a SH-H group (S25) on the 50%S-edge (the reverse reaction path S21 to S25 corresponds to a heterolytic dissociative adsorption of $H_2$ on this edge, which turns out to be easier). Blue line: Free energy at 575K, $pH_2$=10 bar, $pH_2S$=0.1 bar. Red line: Free energy at 675K, $pH_2$=10 bar, $pH_2S$=0.1 bar. Energies of the S21 state are taken as references for each reaction path.

**Figure 8:** Reaction path for the diffusion of H on the 50%S-edge. Blue line: Free energy at 575K, $pH_2$=10 bar, $pH_2S$=0.1 bar. Red line: Free energy at 675K, $pH_2$=10 bar,

pH$_2$S=0.1 bar. Energies of the S21 state (see Fig. 7) are taken as references for each reaction path.

**Figure 9:** Gibbs energy (at T=623K and P(H$_2$)=10 bars) of several states on the M-edge 50%S according to the non-hydrogenated M-edge 50%S (Me50S) in terms of the Sulfur chemical potential. The Me50S_SH_S_SH state corresponds to the S4_bis state, and the state Me50S_SH_HS corresponds to the S4 state.

**Figure 10**: Free energy of the S-edge for different configurations according to the Sulfur chemical potential at T=623K and 10 bar of H$_2$ pressure.

**Figure 11:** a) Phase diagram at 623 K, on the M-edge 50% S, in terms of the H$_2$ relevant HDS conditions (bounded by the brown dashed lines, total pressure between 1 and 200 bar and ratio<10$^{-1}$), corresponding to S1, S6, S4_bis state (see figure 2 and 3). The dashed narrow black line represent the continuation of the limit between the S1 state and the S6 state. The crossed square dot represents the following conditions: p(H$_2$) = 10 bars, p(H$_2$S) = 0.1 bar and Temperature = 623K as indicated by Bollinger et al. [15], [16]. The triangle dot represents the following conditions:  p(H$_2$) = 100 bars, p(H$_2$S) = 0.1 bar and Temperature = 623K. b) Same as 11a) but without including thermal and entropic corrections on surface states energies.

**Figure 12: a)** Phase diagram of the S-edge at 623 K according to the H$_2$ pressure and to the p(H$_2$S)/p(H$_2$) ratio. The state Se100%S+S50%H is represented as figure 5b), and the state Se62%S as figure 5c). The brown dashed lines represents the boundary conditions of HDS, a total pressure of 1 bar, a total pressure of 200 bars and a 10% pH$_2$S/pH$_2$ ratio. The assumed HDS conditions (10 bar of H$_2$ pressure and p(H$_2$S)/(H$_2$) ratio of 0.1) by Bollinger *et al*. [15], [16], are represented by the crossed square dot. b) Same as 12a) but without including thermal and entropic corrections on surface states energies.

**Figure 13**: M-edge and S-edge phase diagrams at 473 K (a-b), 573 K (c-d). As in figure 11 and 12 the dashed brown lines represents the HDS boundary conditions (total pressure between 1 and 200 bars, and maximum pH$_2$S/pH$_2$ ratio 0.2). The assumed HDS conditions (10 bar of H$_2$ pressure and p(H$_2$S)/(H$_2$) ratio of 0.1) by Bollinger *et al*. [15], [16] are represented by the crossed square dots.

**Figure 14**: Activation energies versus reaction energies for H$_2$ and H$_2$S associative desorption processes. A BEPR is brought out for each process. The squared coefficient of correlation is indicated below each regression line equation. All the values come from our present results except from one value (circled point) on each set of values coming from Mo-edge 100%S from N. Dinter *et al*. [3].

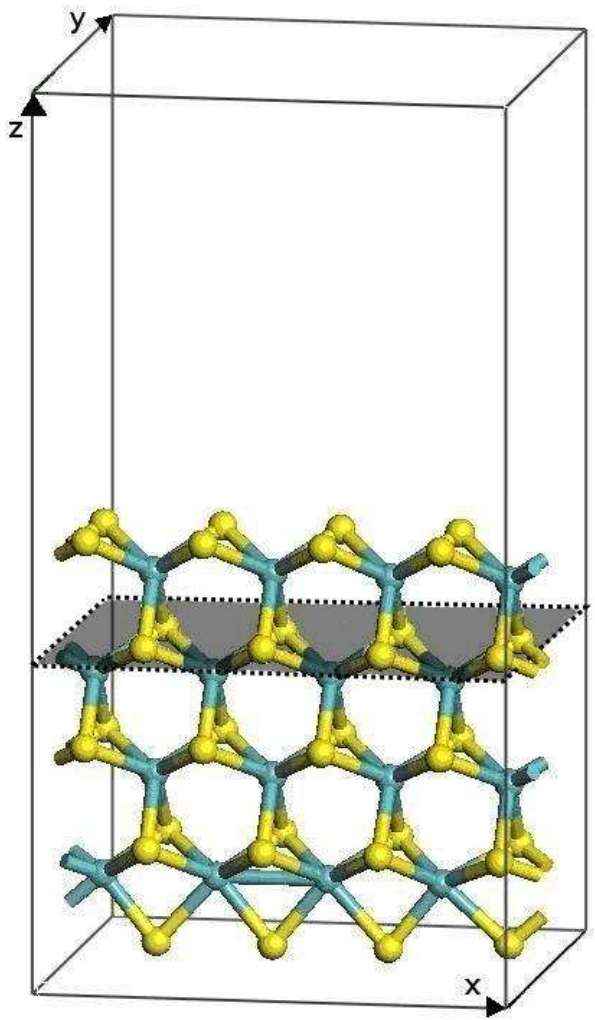

**Figure 1:** A typical periodic slab model for $MoS_2$ S-edge, showing coverage by four $S_2$ bridges. Legend: (color on line) black spheres (turquoise): molybdenum atoms, grey spheres (yellow): sulfur atoms. For normal mode computations, only hydrogen, sulfur, and molybdenum atoms positions above the grey plane were kept free, while the other atoms were frozen.

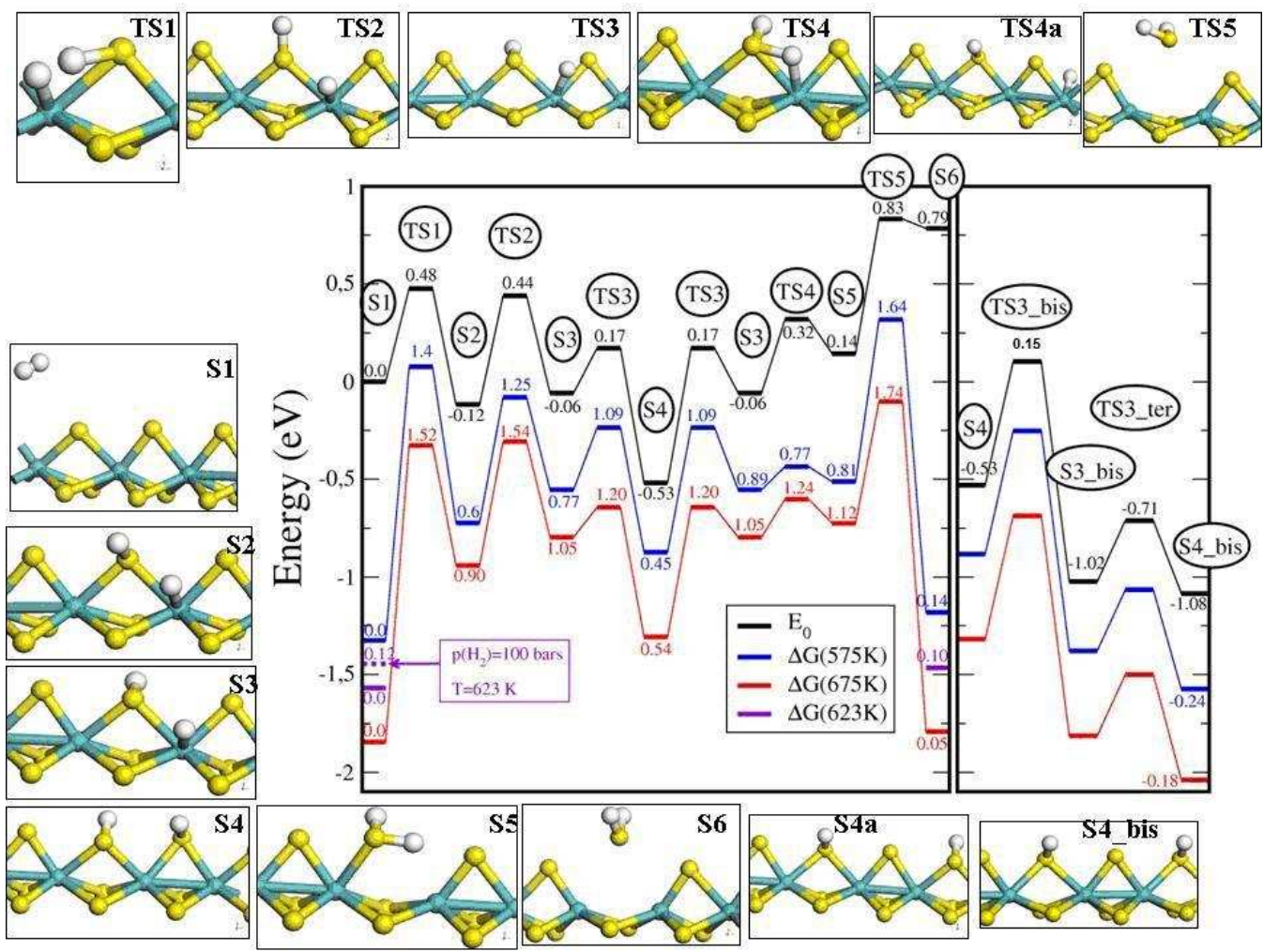

**Figure 2:** Reaction pathway for the creation of one S vacancy on the M-edge 50%S. Blue line: Free energy at 575K, $pH_2=10$ bar, $pH_2S=0.1$ bar ($\Delta\mu_S=-1.01$ eV). Red line: Free energy at 675K, $pH_2=10$ bar, $pH_2S=0.1$ bar ($\Delta\mu_S=-1.13$ eV). Energies of the S1 state are taken as references for each reaction path. On the ordinates axis is reported (purple dash) at 623 K the free energy of the S1 state at 100 bars and 10 bars of $H_2$ pressure and the S6 at 0.1 bar of $H_2S$ pressure. .

**S4_bis**

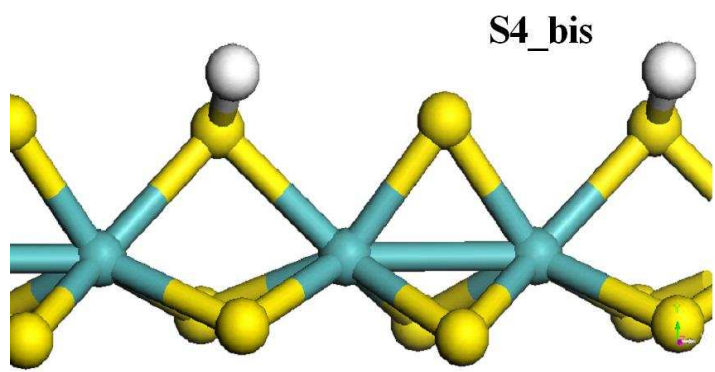

**Figure 3:** Configuration of the most stable state on the M-edge 50%S (-1.09 eV), the S4_bis state, also called Me50S_SH_S_SH.

**4a)**

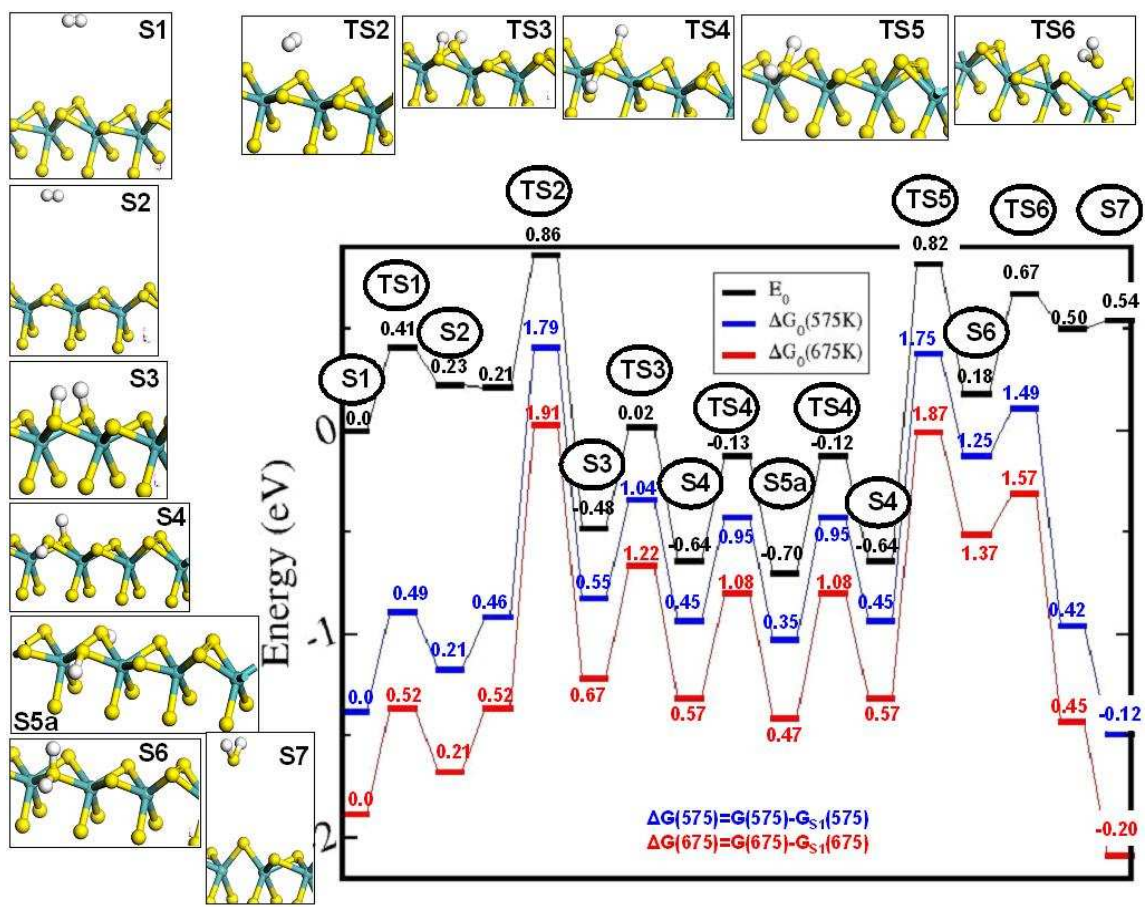

**4b)**

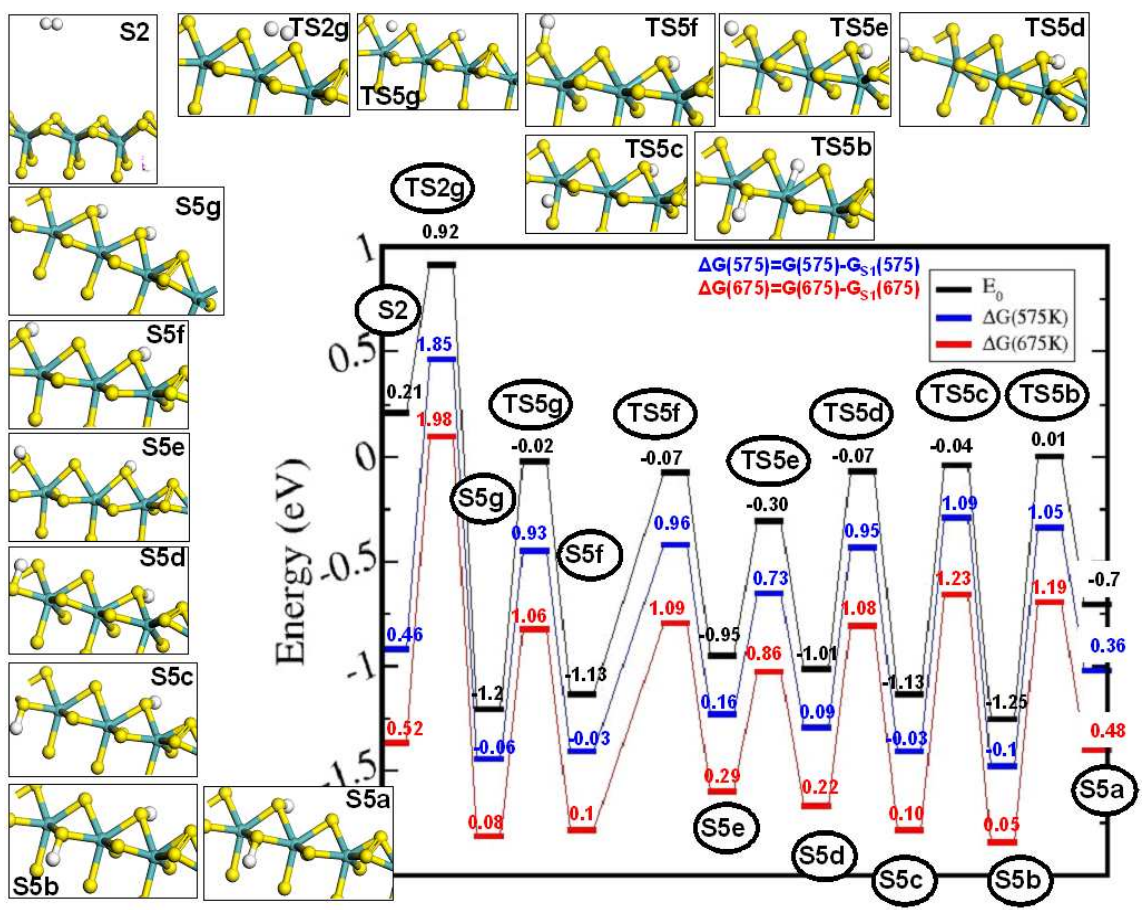

**Figure 4 a):** Reaction pathway for the $H_2S$ desorption process on the S edge 100%S, through $H_2$ adsorption on one bridge. The first step is the separation of the 'dimerized' Sulfur in the bridge. Blue line: Free energy at 575K, p$H_2$=10 bar, p$H_2$S=0.1 bar. Red line: Free energy at 675K, p$H_2$=10 bar, p$H_2$S=0.1 bar. Energies of the S1 state are taken as references for each reaction path. **b):** $H_2$ adsorption on two neighboring bridges and diffusion along the S edge 100%S (S2 and S5a states are equivalent to those on figure 4a). Blue line: Free energy at 575K, p$H_2$=10 bar, p$H_2$S=0.1 bar. Red line: Free energy at 675K, p$H_2$=10 bar, p$H_2$S=0.1 bar. Energies of the S1 state (in figure 4a)) are taken as references for each reaction path. Note that the most stable state at 0K is S5b, while at 675K (red path) it becomes S7.

a)

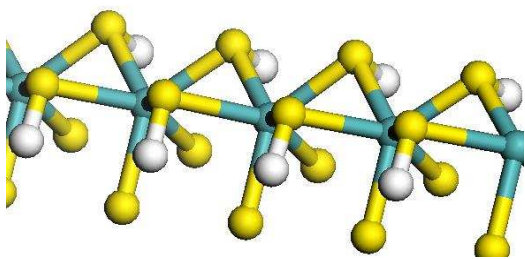

b)

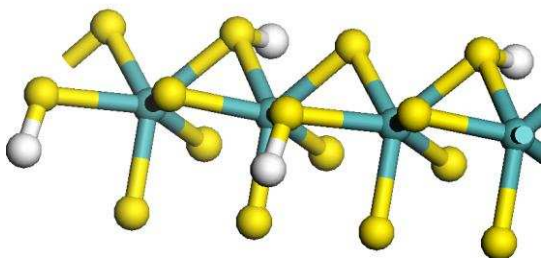

c)

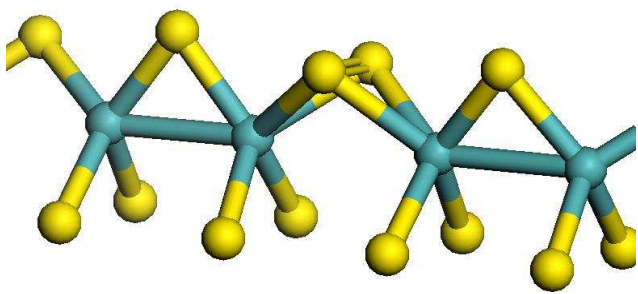

**Figure** 5: a) S-edge 100%S+100%H and b) S-edge 100%S+50%H, respectively 1.3 eV and 2.6 eV lower in energy than S1 state with $H_2$ in gas phase. i.e. -0.325 eV and -1.3 eV adsorption energy per $H_2$ molecule. c) configuration of the S-edge 62%S involved in the phase diagram of figure 11.

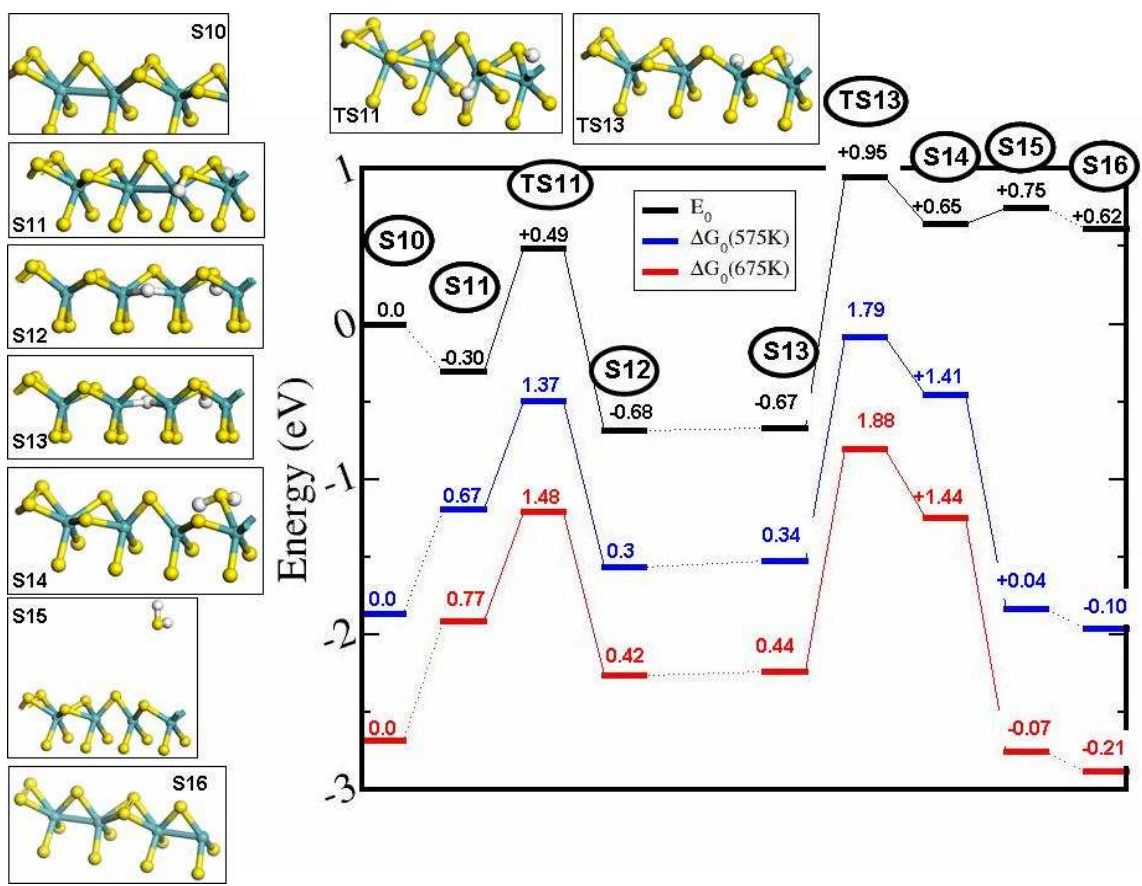

**Figure 6:** Reaction path for $H_2S$ desorption from the 87%S-edge. Blue line: Free energy at 575K, $pH_2$=10 bar, $pH_2S$=0.1 bar. Red line: Free energy at 675K, $pH_2$=10 bar, $pH_2S$=0.1 bar. Each reference energy is the energy of the S10 state for each reaction path. Note that the S15 configuration is less favourable by $\Delta G(T)$ = -0.13 eV (the dependence in temperature in the range of interest is less than 0.01 eV) than S16, where 'dimerized' $S_2$ bridge alternate with S bridges on the edge. (For simplification, $H_2$ molecule in inset S10 and $H_2S$ molecule in insets S16 have been omitted)

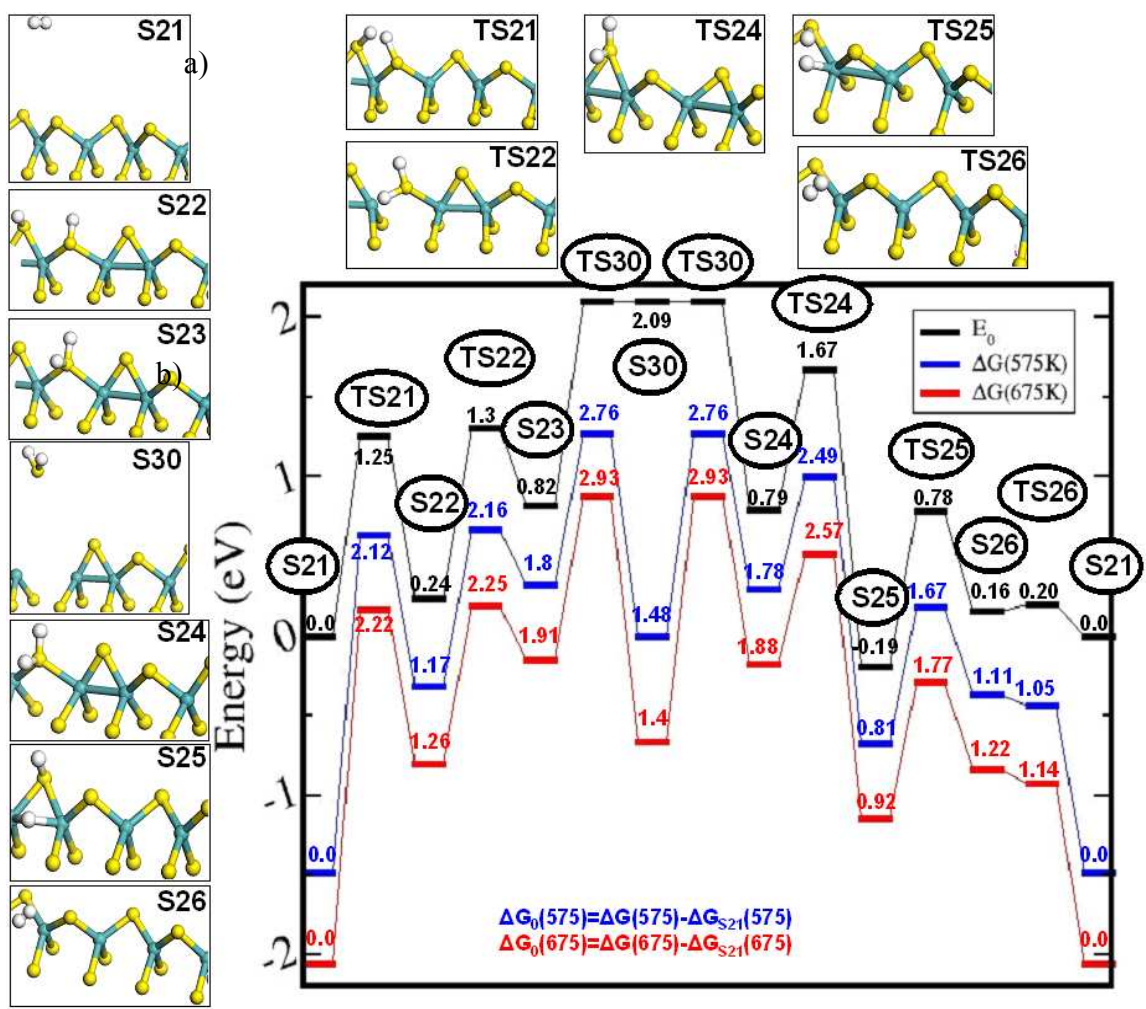

**Figure 7:** Reaction path on the S-edge 50%S for the $H_2$ homolytic dissociative adsorption on two neighboring S bridges (S21 to S22), then $H_2S$ desorption (S23 to S30), followed by $H_2S$ adsorption (S30 to S24), eventually $H_2$ heterolytic associative desorption via a SH-H group (S25) on the 50%S-edge (the reverse reaction path S21 to S25 corresponds to a heterolytic dissociative adsorption of $H_2$ on this edge, which turns out to be easier). Blue line: Free energy at 575K, $pH_2$=10 bar, $pH_2S$=0.1 bar. Red line: Free energy at 675K, $pH_2$=10 bar, $pH_2S$=0.1 bar. Energies of the S21 state are taken as references for each reaction path.

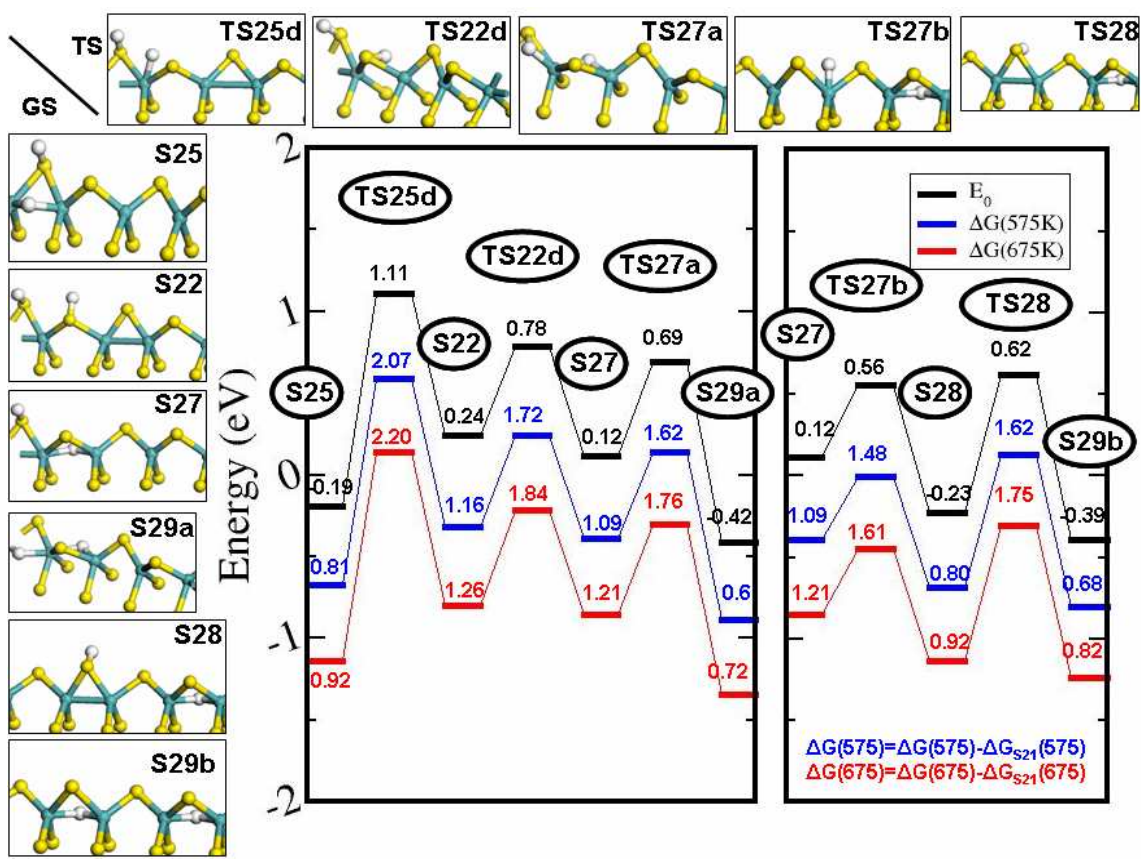

**Figure 8:** Reaction path for the diffusion of H on the 50%S-edge. Blue line: Free energy at 575K, $pH_2$=10 bar, $pH_2S$=0.1 bar. Red line: Free energy at 675K, $pH_2$=10 bar, $pH_2S$=0.1 bar. Energies of the S21 state (see Fig. 7) are taken as references for each reaction path.

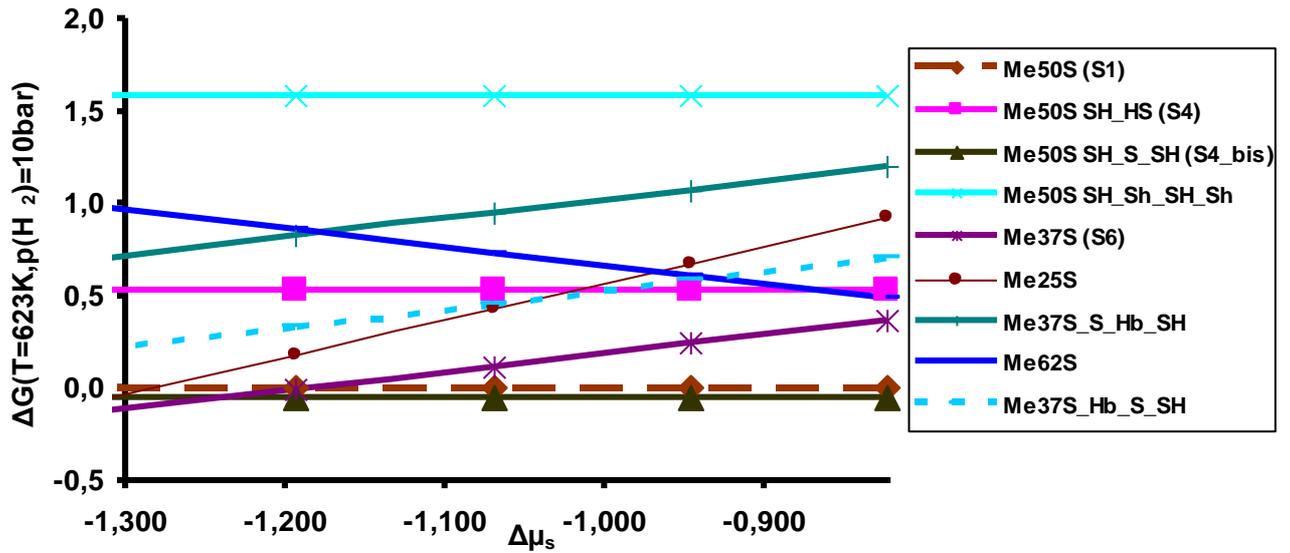

**Figure 9:** Gibbs energy (at T=623K and P(H$_2$)=10 bars) of several states on the M-edge 50%S according to the non-hydrogenated M-edge 50%S (Me50S) in terms of the Sulfur chemical potential. The Me50S_SH_S_SH state corresponds to the S4_bis state, and the state Me50S_SH_HS corresponds to the S4 state.

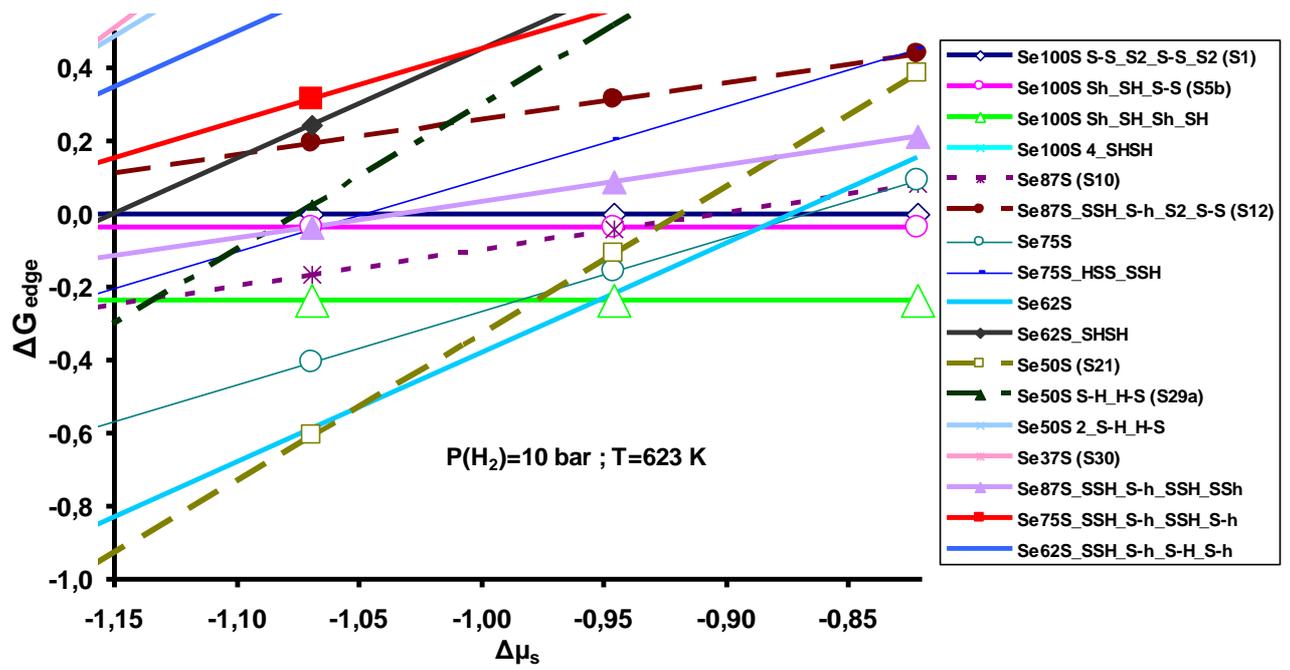

**Figure 10** : Free energy of the S edge for different configurations according to the Sulfur chemical potential at T=623K and 10 bar of H₂ pressure.

**a)**

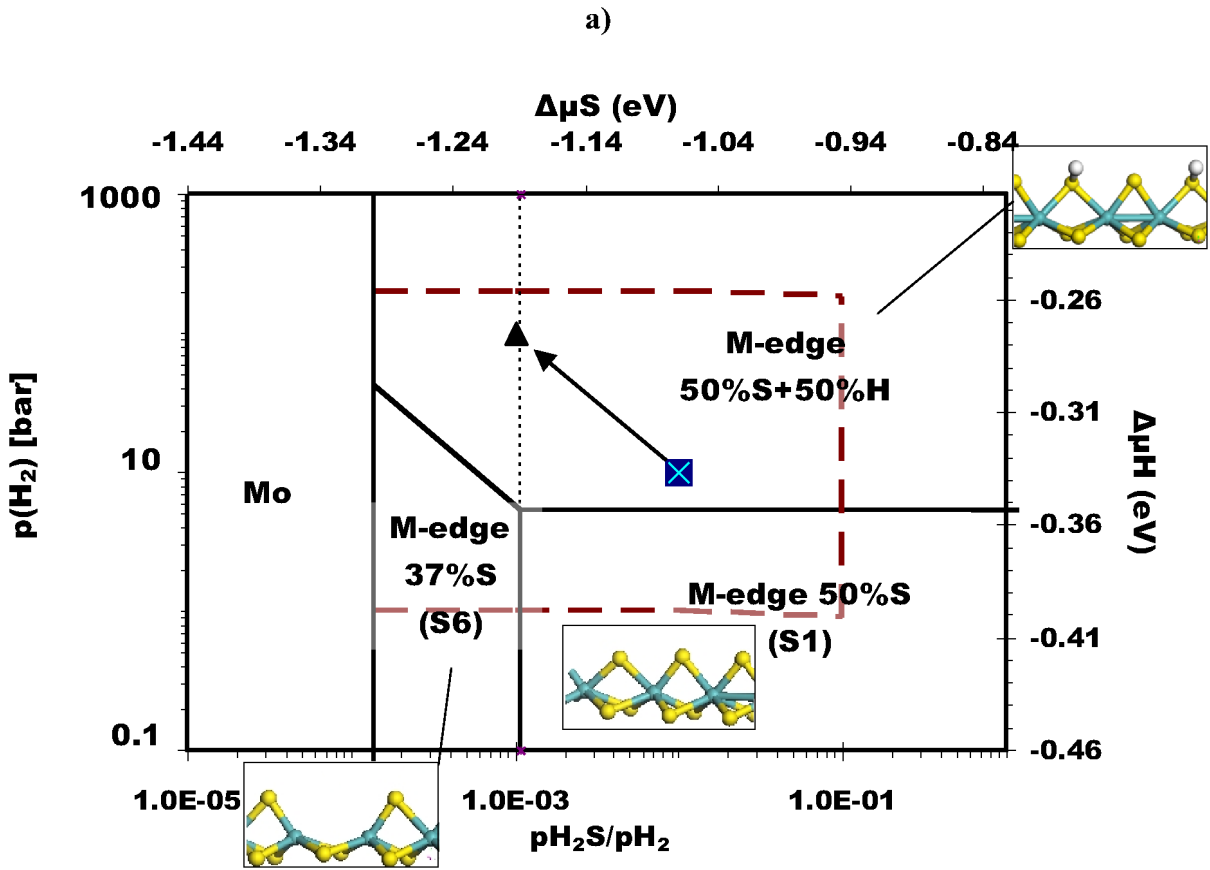

**b)**

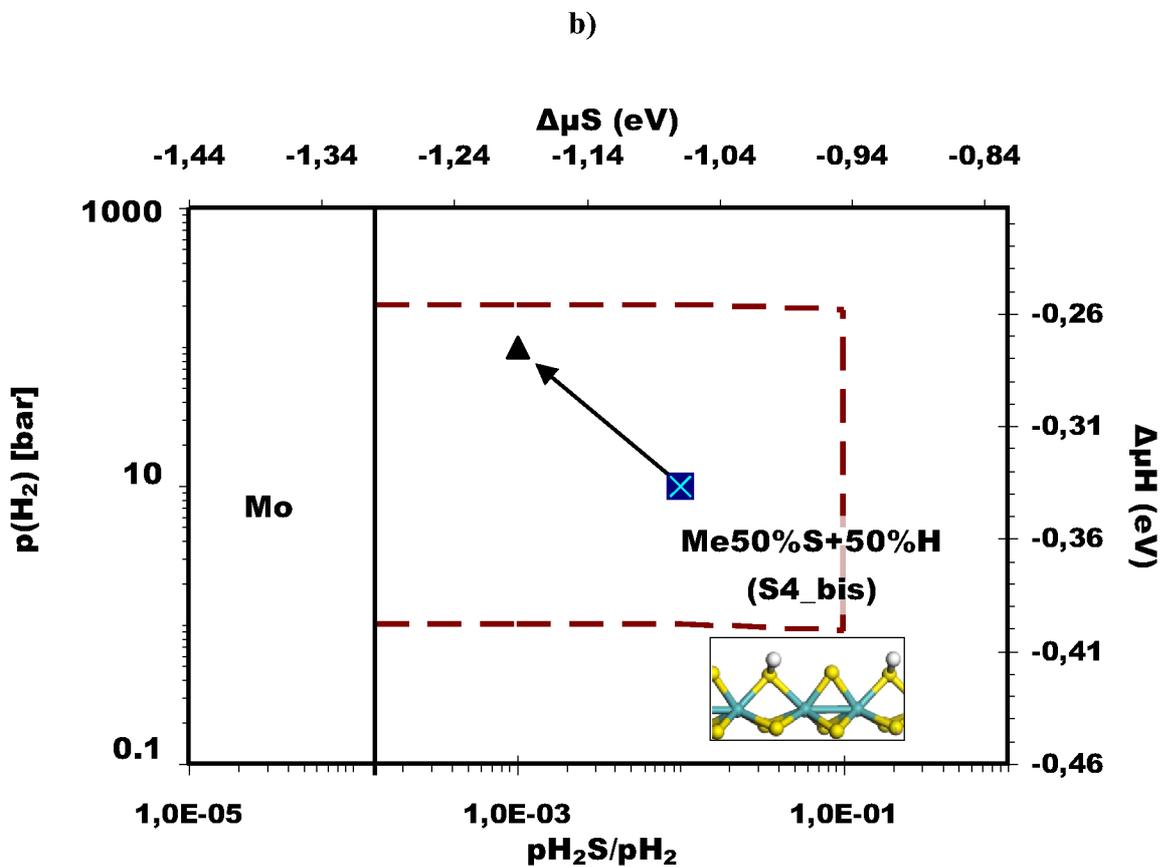

**Figure 11:** a) Phase diagram at 623 K, on the M-edge 50% S, in terms of the $H_2$ relevant HDS conditions (bounded by the brown dashed lines, total pressure between 1 and 200 bar and ratio<$10^{-1}$), corresponding to S1, S6, S4_bis state (see figure 2 and 3). The dashed narrow black line represent the continuation of the limit between the S1 state and the S6 state. The crossed square dot represents the following conditions: p($H_2$) = 10 bars, p($H_2S$) = 0.1 bar and Temperature = 623K as indicated by Bollinger et al. [15], [16]. The triangle dot represents the following conditions:  p($H_2$) = 100 bars, p($H_2S$) = 0.1 bar and Temperature = 623K. b) Same as 11a) but without including thermal and entropic corrections on surface states energies.

a)

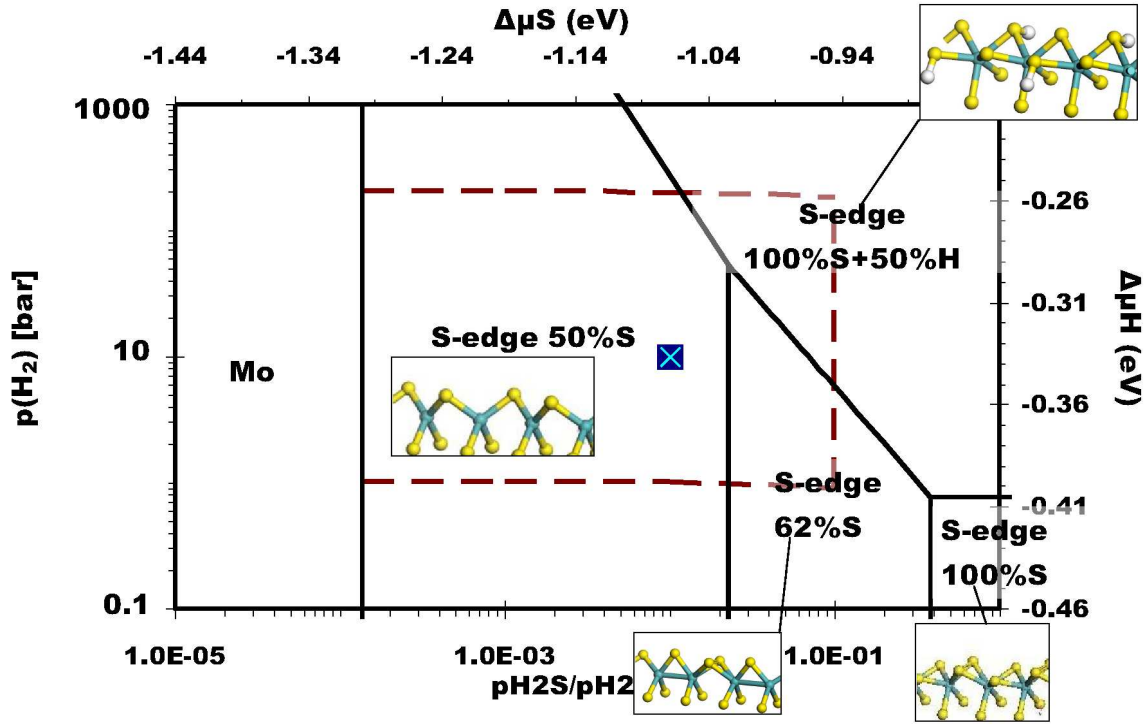

b)

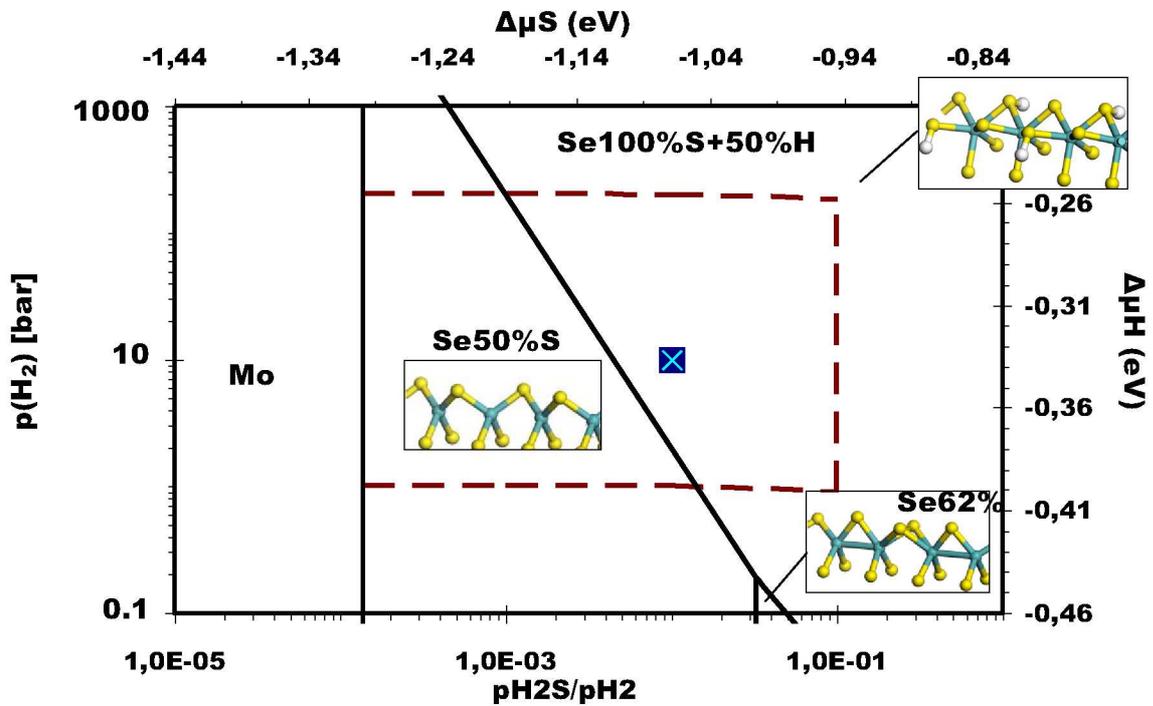

**Figure 12: a)** Phase diagram of the S-edge at 623 K according to the $H_2$ pressure and to the $p(H_2S)/p(H_2)$ ratio. The state Se100%S+S50%H is represented as figure 5b), and the state Se62%S as figure 5c). The brown dashed lines represents the boundary conditions of HDS, a total pressure of 1 bar, a total pressure of 200 bars and a 10% $pH_2S/pH_2$ ratio. The assumed HDS conditions (10 bar of $H_2$ pressure and $p(H_2S)/(H_2)$ ratio of 0.1) by Bollinger *et al.* [15], [16], are represented by the crossed square dot. **b)** Same as 12a) but without including thermal and entropic corrections on surface states energies.

a)

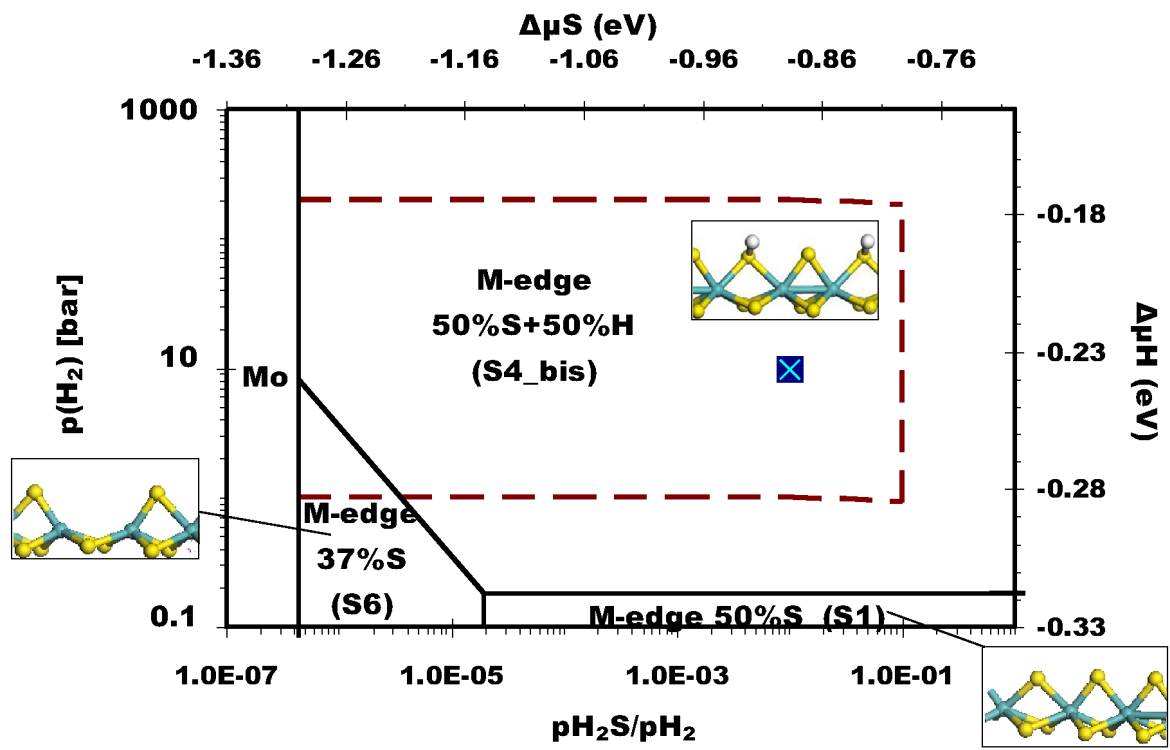

b)

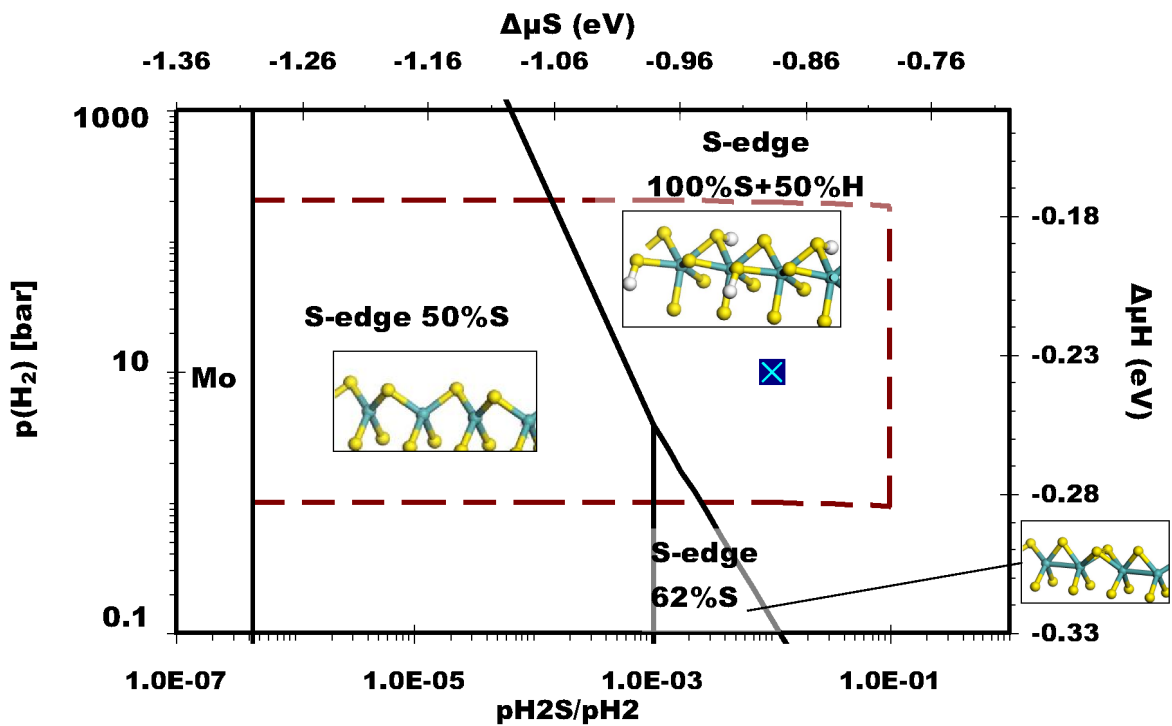

c)

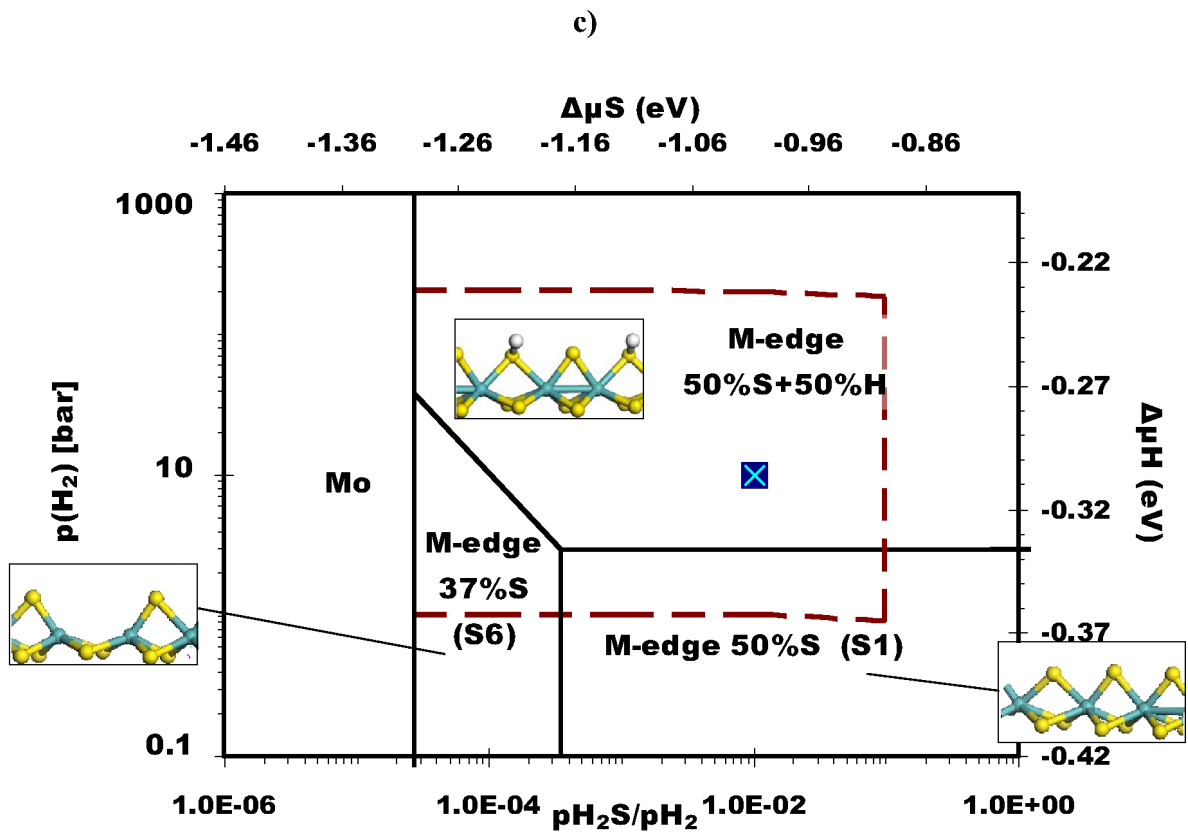

d)

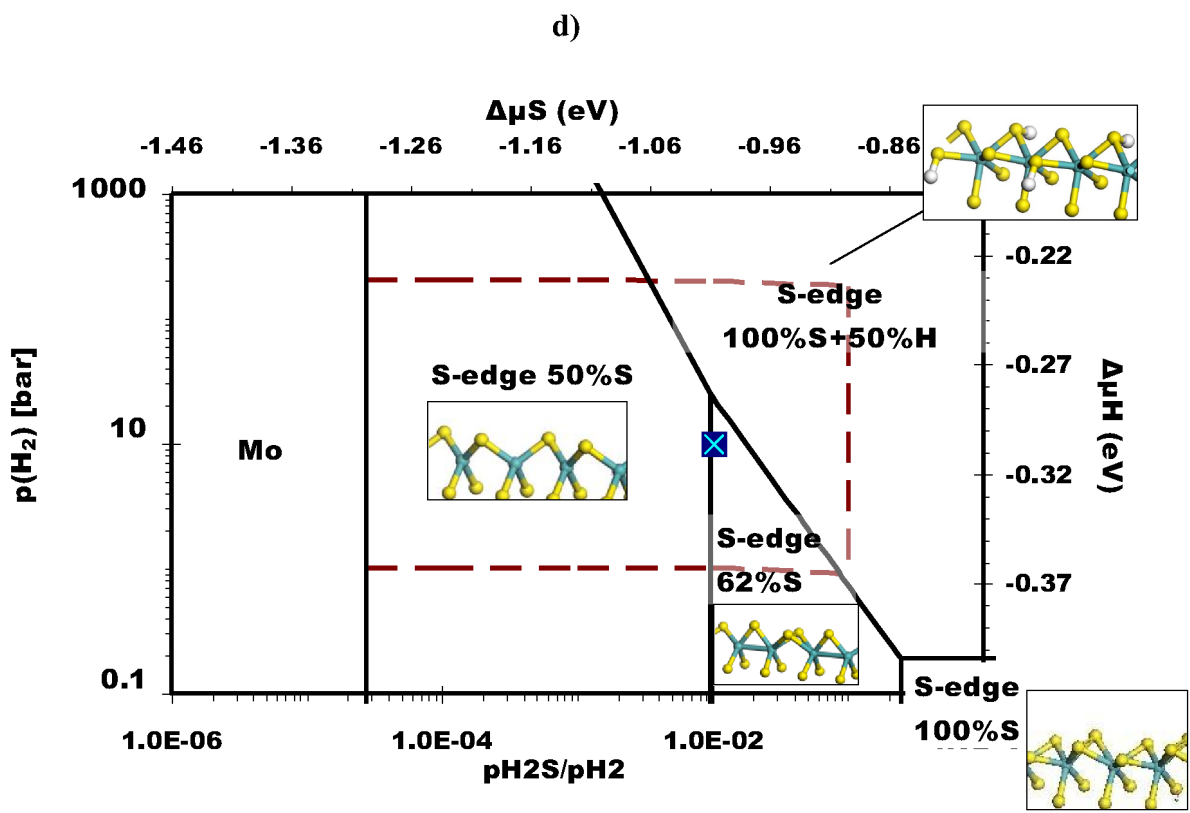

**Figure 13**: M-edge and S-edge phase diagrams at 473 K (a-b), 573 K (c-d). As in figure 11 and 12 the dashed brown lines represents the HDS boundary conditions (total pressure between 1 and 200 bars, and maximum $pH_2S/pH_2$ ratio 0.2). The assumed HDS conditions (10 bar of $H_2$ pressure and $p(H_2S)/(H_2)$ ratio of 0.1) by Bollinger *et al.* [15], [16] are represented by the crossed square dots.

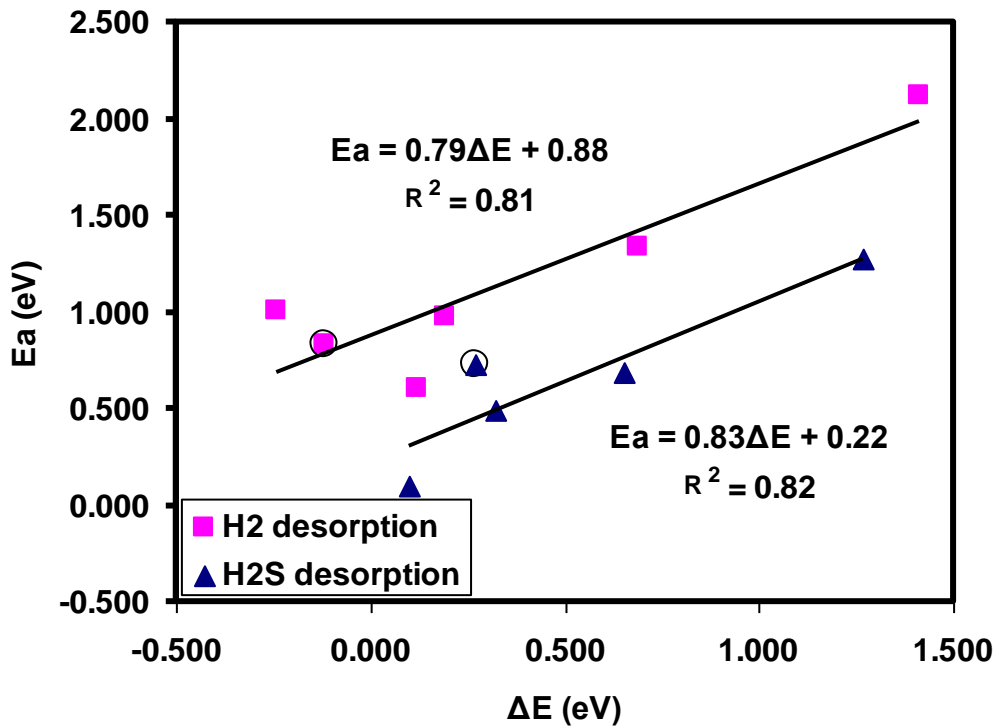

**Figure 14**: Activation energies versus reaction energies for $H_2$ and $H_2S$ associative desorption processes. A BEPR is brought out for each process. The squared coefficient of correlation is indicated below each regression line equation. All the values come from our present results except from one value (circled point) on each set of values coming from Mo-edge 100%S from N. Dinter *et al.* [3].

# References

[1] P. Raybaud, Appl. Catal. A: Gen., 322 (2007) 76.

[2] P. G. Moses, B. Hinnemann, H. Topsøe, J.K. Nørskov, J. Catal., 268 (2009) 201.

[3] N. Dinter, M. Rusanen, P. Raybaud, S. Kasztelan, P. da Silva, H. Toulhoat, J. Catal. 267 (2009) 67

[4] Topsoe, H., Clausen, B.S., Catal. Rev. Sci. Eng., 26 (1984) 395.

[5] S. Kasztelan, H. Toulhoat, J. Grimblot, J.P. Bonnelle, Appl. Catal., 13 (1984) 127.

[6] A. Travert, H. Nakamura, R. Van Santen, S. Cristol, J. Paul, E. Payen, J.A.C.S., 124 (2002) 7084

[7] J.F. Paul, E. Payen, J. Phys. Chem.B 107 (2003) 4057

[8] M. Sun, A.E. Nelson, J. Adjaye, Catalysis Today 105 (2005) 36.

[9] P. G. Moses, B. Hinnemann, H. Topsøe, J.K. Nørskov, J. Catal. 248 (2007) 188

[10] R.J.H. Voorhoeve, J.C.M. Stuiver, J. Catal., 23 (1971) 243.

[11] A. Vambeke, L. Jalowiecki, J. Grimblot, J.P. Bonnelle, J. Catal., 109 (1988) 320.

[12] L.S. Byskov, J.K. Nørskov, B.S. Clausen, H. Topsøe, J. Catal., 187 (1999) 109.

[13] S. Cristol, J.F. Paul, E. Payen, D. Bougeard, S. Clemendot, F. Hutschka,J. Phys. Chem. B, 106 (2002) 5659

[14] M. Sun, A.E. Nelson, J. Adjaye, J. Catal. 233 (2005) 411

[15] M.V. Bollinger, K.W. Jacobsen, J.K. Nørskov, Phys. Rev. B, 67 (2003) 85410

[16] Lauritsen J.V., Bollinger M.V., Lægsgaard E., Jacobsen K.W., Nørskov J.K., Clausen B.S., Topsøe H., Besenbacher F., J. Catal. 221 (2004) 510.

[17] W. Kohn, L.J. Sham, Phys. Rev. 140 (1965) A1133.

[18] Kresse G., Furthmüller J., Comput. Mater. Sci. 6 (1996) 15.

[19] J.P. Perdew, J.A. Chevary, S.H. Vosko, K.A. Jackson, M.R. Pederson, D.J. Singh, C. Fiolhais, Phys. Rev. B, 46 (1992) 6671.

[20] J.P. Perdew, Y. Wang, Phys. Rev. B, 45 (1992) 13244

[21] G. Kresse, D. Joubert, Phys. Rev. B, 59 (1999) 1758.

[22] H.J. Monkhorst, J.D. Pack, Phys. Rev. B 13 (1976) 5188.

[23] H. Jónsson, G. Mills, and K. W. Jacobsen, *Classical and Quantum Dynamics in Condensed Phase Simulations*, edited by B. J. Berne, G. Ciccotti, and D. F. Coker (World Scientific, Singapore, 1998), p. 385.

[24] G. Henkelman, B.P. Uberuaga, and H. Jónsson, J. Chem. Phys. 113 (2000), 9901

[25] G. Henkelman, H. Jónsson, J. Phys. Chem, 111 (1999) 7010.

[26] R. F. W. Bader, *Atoms in Molecules - A Quantum Theory*, Oxford University Press, Oxford, 1990.

[27] G. Henkelman, A. Arnaldsson, and H. Jónsson, Comput. Mater. Sci. 36 (2006) 254.

[28] E. Sanville, S. D. Kenny, R. Smith, and G. Henkelman, J. Comp. Chem. 28 (2007) 899.

[29] W. Tang, E. Sanville, and G. Henkelman, J. Phys.: Comp. Mater. 21 (2009) 084204.

[30] R.A. van Santen, J.W. Niemantsverdriet, *Chemical Kinetics and Catalysis*, Plenum Press, New York, 1995.

[31] P. Raybaud, J. Hafner, G. Kresse, S. Kasztelan, H. Toulhoat, J. Catal. 189 (2000) 129.

[32] P. Raybaud, J. Hafner, G. Kresse and H. Toulhoat, Phys. Rev. Lett., 80 (1998) 1481.

[33] P. Raybaud, H. Toulhoat, J. Hafner, G. Kresse, *Proc. 2nd Intern. Symposium on Hydrotreatment and Hydrocracking of Oil Fractions*, ed. by B. Delmond, J. F. Froment, and P. Grange, Studies in Surf. Sci. and Catal. 127 (1999) 309-317.

[34] B. Hinnemann, J.K. Nørskov, H. Topsøe, J. Phys. Chem. B 109 (2005) 2245

[35] R. A. Van Santen, M. Neurock, S.G. Shetty, Chem. Rev., 110-4 (2010) 2005.

[36] H. Schweiger, P. Raybaud, G. Kresse, H. Toulhoat, J. Catal. 207 (2002) 76.

[37] E. Krebs, B. Silvi, P. Raybaud, Catal. Today 130 (2008) 160–169

[38] J. Polz, H. Zeilinger, B. Müller, & H. Knözinger, J. Catal., 120 (1989) 22 - 28

[39] P. Sundberg. R.B. Moyes, J. Tomkinson, Bull. Soc. Chim. Belg., 100 (1991) 967.

[40] J.A. Spirko, M.L. Neiman, A.M. Oelker, K. Klier, Surf. Sci., 572 (2004) 191.

[41] T.E. Burrow, N.J. Lazarowych, R.H. Morris, J.Lane, R.L. Richards, Polyhedron, 8 (1989) 1701.

[42] N. Dinter, M. Rusanen, P. Raybaud, S. Kasztelan, P. da Silva, H. Toulhoat, J. Catal. 275 (2010) 117.

[43] J.F. Paul, S. Cristol, E. Payen, Catal. Today, 130 (2008) 139.

[44] See for instance http://en.wikipedia.org/wiki/Equation_of_state

[45] S. Kasztelan and D. Guillaume, Ind. Eng. Chem. Res., 33, (1994), 203.